\documentclass[a4paper, 11pt] {article}
\pdfoutput=1
\usepackage{graphicx}
\usepackage{jheppub}
\usepackage{amsmath}
\usepackage{amsthm}
\usepackage{amsfonts}
\usepackage{amssymb}
\usepackage{textcomp}
\usepackage{footmisc}
\usepackage{bm}
\usepackage{color}
\usepackage{verbatim}
\usepackage{subcaption}
\usepackage[abs]{overpic}

\theoremstyle{definition}

\definecolor{orange}{rgb}{1,0.5,0}
\definecolor{col1}{RGB}{153, 52, 121}
\definecolor{dgreen}{rgb}{0,0.55,0}
\definecolor{pink}{rgb}{1,0.08,0.58}

\usepackage{amsfonts,amsmath,amssymb}

\begin{document}

\title{
Evidence for violations of Weak Cosmic Censorship in black hole collisions in higher dimensions}

 \author[a]{Tomas Andrade,}

 \author[b]{Pau Figueras}
 
 \author[c,d]{Ulrich Sperhake}

\affiliation[a]{
Departament de F{\'\i}sica Qu\`antica i Astrof\'{\i}sica, Institut de
Ci\`encies del Cosmos, Universitat de
Barcelona, Mart\'{\i} i Franqu\`es 1, 08028 Barcelona, Spain}

\affiliation[b]{School of Mathematical Sciences, Queen Mary University of London, Mile End Road, London E1 4NS, United Kingdom
}%

\affiliation[c]{DAMTP, Centre for Mathematical Sciences, University of Cambridge, Wilberforce Road, Cambridge CB3 0WA, UK} 
\affiliation[d]{California Institute of Technology, Pasadena, California 91125, USA}

\emailAdd{tandrade@icc.ub.edu}
\emailAdd{p.figueras@qmul.ac.uk}
\emailAdd{U.Sperhake@damtp.cam.ac.uk}


\abstract{We study collisions of boosted rotating black holes in $D=6$ and 7 spacetime dimensions  with a non-zero impact parameter. 
We find that there exists an open set of initial conditions such that the intermediate state of the collision is a black hole with a dumbbell-like horizon which is unstable to a local Gregory-Laflamme-type instability. We are able to provide convincing numerical evidence that the evolution of such an instability leads to a pinch off of the horizon in finite asymptotic time thus forming a naked singularity, as in similar unstable black holes. Since the black holes in the initial state are stable, this is the first genuinely generic evidence for the violation of the Weak Cosmic Censorship Conjecture in higher dimensional asymptotically flat spacetimes.}

\maketitle

\section{Introduction}
\label{sec:Intro}

General Relativity (GR) governs the large scale structure of our Universe. 
Despite its enormous success, the fact that singularities can be generically formed in dynamical settings \cite{Hawking:1969sw} in principle calls into question its predictive power. 
Motivated by this, Penrose postulated the ``Weak Cosmic Censorship (WCC) Conjecture'' \cite{Penrose:1969pc}, which proposes that singularities 
formed in dynamical evolution must be hidden inside black hole (BH) horizons. 
This constitutes a key link between the singularity theorems and the ubiquitousness of astrophysical
black holes, for which Penrose was recently awarded the Nobel prize.
Through active investigation in over half a century, counter-examples to WCC
in settings beyond astrophysics have been found, which can be regarded as a possibility to access the quantum regime of gravity, at least theoretically. 
They fall into two major categories: i) critical collapse, which involves fine tuning of initial data such that a zero mass singularity is formed in $4D$ and above \cite{Choptuik:1992jv}; 
ii) death by fragmentation, which results from elongated horizons becoming unstable and eventually pinching off in finite time. 
The latter category will be the main focus of this article. 

The instability which drives the initial stages of this phenomenon was famously discovered by Gregory and Laflamme (GL) \cite{Gregory:1993vy}, who showed, in $5D$ and above, that thin enough black strings (and $p$-branes in general) have exponentially growing linear modes that trigger spatial modulation along the horizon, in a way that resembles the Rayleigh instability in fluids \cite{Eggers:1993aa, Eggers:1997aa}.  
Later on, \cite{Lehner:2010pn,Lehner:2011wc} used numerical relativity to follow the unstable black string into the non-linear regime and 
 provided convincing evidence in favour of the pinching off of the horizon in finite time. Moreover, the authors 
observed the formation of BH satellites joined by thin necks organized in a fractal structure. 
It is by now understood that not only strings, but all elongated horizons which are locally string (or brane)-like and thin enough, 
present these unstable modes, e.g., black rings \cite{Santos:2015iua}, and ultra-spinning BHs
\cite{Emparan:2003sy, Dias:2009iu, Dias:2010eu, Shibata:2009ad, Shibata:2010wz, Dias:2014eua}.
The full non-linear evolution of such unstable configurations has been studied in \cite{Figueras:2015hkb, Figueras:2017zwa, Bantilan:2019bvf}. 
In Refs. \cite{Lehner:2010pn,Lehner:2011wc, Figueras:2015hkb, Figueras:2017zwa, Bantilan:2019bvf} the quantity that is used to track the shape of the horizon is in fact the apparent horizon. This is justified in view of the quasi-stationary evolution of most of the geometry after the first generation of satellites has formed, which in turn implies that the event horizon is well approximated by the apparent horizon almost everywhere. Even though none of these studies addresses the pinch off of the event horizon and the completeness of future null infinity, it is clear that during the evolution of the GL instability, arbitrarily large curvatures become visible for asymptotic observers in finite time, thus violating at least the spirit of the WCC.

All of the above setups have in common that the starting point is an unstable configuration. Therefore, one may wonder if the so-obtained singularities are a truly generic feature of the evolution in GR. 
Along these lines, a new and truly generic mechanism for violation of WCC has recently been proposed \cite{Andrade:2018yqu, Andrade:2019edf}: the collision of BHs in higher dimensions. The authors  of these papers noted that if the total angular momentum is large enough, an elongated horizon should form as an intermediate, long-lived, state after the collision. 
After this point, the horizon could undergo the breaking described above, depending on the competition between radiation of mass and angular momentum, and the GL instability. The former tends to make the horizon rounder and shorter, thus taming the effect of GL.
Since radiation is exponentially suppressed with $D$ and the growth rates of the GL instability remain relatively constant, this mechanism is guaranteed to work if $D$ is sufficiently large. 
On the other hand, this behaviour is absent in head-on collisions in four or higher  dimensions \cite{Sperhake:2008ga,Sperhake:2019oaw}, 
and in grazing BH collisions in 4D \cite{Sperhake:2009jz}, none of which exhibit elongated horizons or any signs of censorship violation even when colliding above 90\,\% of the speed of light. 
Uncovering exactly when and how the proposed phenomenon takes place, requires fully-fledged numerical GR techniques. 
We note that bar mode instabilities of rotating black holes are known to exist in 6$D$ and above \cite{Shibata:2010wz,Dias:2014eua}, so  elongated horizons are more likely to exist in $D\ge 6$.
For a sufficiently large initial angular momentum, these unstable `bar-shaped' black holes develop a local GL instability.  During the evolution of this instability, only about $5\%$ of the total mass of the system is radiated, which is insufficient to prevent the formation of a naked singularity in finite time \cite{Bantilan:2019bvf}.
Furthermore, Ref.~\cite{Sperhake:2019oaw} reports that for $v \leq 0.6$, the radiation emitted in head-on collisions is below $1\%$.  All of this considered, we envision that intermediate velocities, relatively high intrinsic spins and impact parameters, and $D \geq 6$ constitute a favourable region for WCC violation in parameter space.

\begin{figure}[t!]
\centering
\includegraphics[scale=0.75]{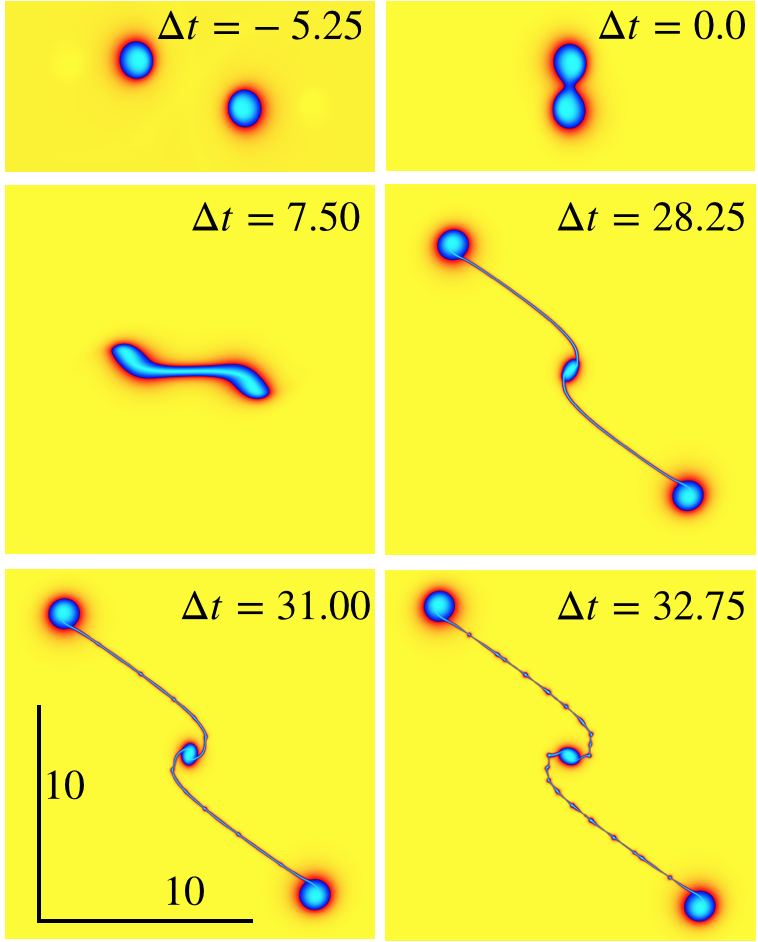}
\caption{%
Snapshots of the 7$D$ evolution with collision velocity $v= 0.5$, dimensionless spin $a = 0.7$, and impact parameter $b = 2.2$, at times $\Delta t = t - t_\text{merger}$. They correspond to: initial condition, merger,  bar formation, first generation, second generation, latest stage. We show a spatial length of 10 units on the bottom left panel.}
\label{fig:Bin7d}
\end{figure} 

Here we undertake the task of numerically simulating the grazing collision of $6D$ and $7D$ stable, single-spinning BHs with a finite impact parameter. This initial setting is closely analogous to that of a BH binary merger in $4D$, but the spacetime dimensionality gives rise to striking differences in the evolution as we illustrate in Fig.~\ref{fig:Bin7d} for a specific collision in $7D$.
All our simulations start with two separate equal-mass BHs with spin parameter $a$, collision velocity $v$ and impact parameter $b$; cf.~the first (upper left)
panel in Fig.~\ref{fig:Bin7d}. Over a significant portion of the parameter
space, we find that these configurations to form an elongated horizon shortly after
merger (second and third panels in the figure).
The resulting ``neck'' grows ever thinner (fourth panel) and breaks through the further evolution of multiple generations of local GL instabilities along the individual string segments (fifth and sixth panel). This process does not require any fine tuning.
Moreover, we observe that the late time dynamics resemble the end point of the GL instabilities of black strings and ultra-spinning BHs of \cite{Figueras:2015hkb, Figueras:2017zwa, Bantilan:2019bvf}. In particular, we observe the formation of two well-defined generations of satellites, and the beginning of a third one. Each generation consists of spherical blobs joined by thin tubes, which can be approximated by quasi-stationary
BHs and black strings, respectively. The last stage to which we have access to consists of a central rotating black object, joined by a thin horizon to two non-spinning blobs that move away from each other, see Fig \ref{fig:Bin7d}; the animation for the full time range covered in our numerics is available at \cite{YoutubeVideo1}. The thin tubes in the middle are stretching due to the motion of the outer blobs, and develop their own satellites due to their own local GL instabilities.  Hence, it is clear that the process will continue until the pinch off, which will take place in finite asymptotic time. 

We extract the gravitational waves produced in the collision, finding that the waveforms exhibit two salient peaks. The first one is simply due to the initial merger of the BHs. The second peak produces a stronger signal, and we attribute it to the sudden change of angular velocity and shape of the elongated horizon. 
Employing a toy model of a $D>4$ dimensional binary which undergoes a qualitatively similar change in angular velocity and separation, we extract the radiated energy and observe a similar peak; see Appendix \ref{app:gw_toymodel}. 
We find that the total radiated energy is strongly suppressed (it amounts to less than $1 \%$ in 7D), confirming the hypothesis that there is no mechanism to stop the formation of a naked singularity in finite asymptotic time.

The organization of the rest of the paper is as follows: In Section \ref{sec:num_methods} we describe the numerical methods that we have used in this paper. Section \ref{sec:AHandchi} discussed the relation between the apparent horizon and the contours of the conformal factor $\chi$ in the moving punctures gauge that we use. We present our main results for BH collisions in $6D$ and $7D$ in Section \ref{sec:results}.
We close with a discussion of our results and possible directions for future research in Section \ref{sec:discussion}. We have relegated some technical results to several appendices.  

\section{Numerical methods}
\label{sec:num_methods}
We solve the vacuum Einstein equations in $D = 6, 7$ using the CCZ4 formulation 
as described in \cite{Alic:2011gg, Weyhausen:2011cg}. We restrict ourselves to dynamics in 3+1 dimensions by imposing an 
$SO(D-3)$ symmetry by means of the modified cartoon method \cite{Shibata:2010wz, Pretorius:2004jg, Cook:2016soy}.
The conventions for the damping parameters $\kappa_1$ and $\kappa_2$, and choice of gauge evolution are those of Ref.~\cite{Bantilan:2019bvf}. 
We construct the initial data by superimposing two boosted Myers-Perry (MP) BHs \cite{Myers:1986un}
written in Kerr-Schild (KS) coordinates, as described in Ref.~\cite{Sperhake:2006cy}. This yields a mild violation of the 
constraints on the initial data surface, but in higher dimensions the effect is small for sufficiently separated initial black holes; 
see Appendix \ref{app:ID} for details.
Henceforth, we split the coordinates as $t,x,y,z, w_i$ where $\partial_{w_i}$ are ``cartoon'' directions in the language of \cite{Cook:2016soy}, 
and choose a single spin aligned with the $z-$axis. 

Without loss of generality, we place one of the BHs at $(x_0, y_0, 0)$ and give it
an initial velocity along the $x-$direction, which we denote by $v$. 
We ease the tracking of the dynamics by choosing symmetric data in which the second 
BH is positioned at $(- x_0, - y_0, 0)$ with speed $(-v,0,0)$, which gives an
orbital angular momentum pointing in the $z-$direction. 
Both black holes have intrinsic spin along the $z-$direction, with spin parameter $a$. 
We emphasise that in our setting the orbital and intrinsic angular momenta are aligned, which allows us to achieve higher values of the total angular momentum. The impact parameter is controlled by $y_0$. Recall that in $D>4$ 
there is no inspiral phase in binaries \cite{Cook:2018fxg}, which means that, generically, BHs interact only if $y_0$ is sufficiently
small. 
While $x_0$ does not play a fundamental role, we need to choose it such that the initial BHs are sufficiently far apart so that the constraint violations on the initial surface as a result of simply superposing two MP BHs are small. 
In addition, having a sufficiently large $x_0$ allows us to decouple the initial gauge dynamics from the collision; our gauge
conditions result in an adjustment phase lasting about $\sim 3\,\mu^{1/(D-3)}$.

We implement the numerical scheme in \texttt{GRChombo} \cite{Clough:2015sqa, chombo-design-doc}, using up to twelve levels of refinement with a 2:1 refinement ratio.  We add the extra levels  to better resolve the spatial gradients thus keeping the regions of constraint violations far away from the $\chi = 0.6$ surface, which we use as a proxy for the apparent horizon (AH).   Here  $\chi=(\det\gamma)^{-\frac{1}{3}}$ is the conformal factor, where $\gamma_{ij}$ is the induced metric on the spatial slices. Recall that $\chi$ is one of the evolution variables in the CCZ4 formulation of the Einstein equations; we discuss in detail the relation between the level sets of this variable and the AH  in the next section. 
We have checked the robustness of our results by introducing refinement levels at different times, provided that the extra levels are added whilst the region with constraint violations is sufficiently small.  Following \cite{Figueras:2015hkb,Figueras:2017zwa,Bantilan:2019bvf}, we apply artificial diffusion in the region $\chi\leq0.075$ to keep the constraint violations under control. 
Our computational domain, boundary conditions and time evolution scheme are described in \cite{Bantilan:2019bvf}. Finally, we extract the gravitational waves produced in the collisions following Ref.~\cite{Bantilan:2019bvf}; see Appendix \ref{app:gw7d} for details on the wave extraction procedure.

\section{Apparent horizons and contours of $\chi$}%
\label{sec:AHandchi}

Our main observables of interest are the location of the AH and the Kretschmann curvature scalar.  In our setting, tracking the AH from first principles entails
solving a non-linear PDE in two variables at a given instant of time.
Shortly after the collision of the two BHs, the common AH evolves into a non-star-shaped surface, with increasingly complex structure as shown in 
Figs.~\ref{fig:Bin7d}, \ref{fig:Bin6d-1}, \ref{fig:Bin6d-2} there exist as yet no algorithms to find the AH for such surfaces.
%

In this paper, we bypass this difficulty as follows. Reference \cite{Bantilan:2019bvf} provides convincing evidence in a variety of scenarios that, in the standard moving punctures gauge, the location of the AH matches within a few percent with a contour of the conformal factor $\chi$. 
While there is some dependence on the space-time dimension $D$, this has been 
observed to be mild \cite{Figueras:2015hkb,Figueras:2017zwa,Bantilan:2019bvf}. Motivated by this, below we will provide quantitative evidence that, for the non-linear evolution of the GL instability in the cases discussed in this paper, the $\chi = 0.6$ contour should accurately track the location of the ``apparent horizon''. In particular, this contour should capture the essential features of the AH's shape. Other contours, such as $\chi=0.5$ or $\chi=0.4$, would be comparatively informative. This insensitivity to the specific value of the contour of $\chi$  as long as it is sufficiently close to the AH, is due to the steep gradients near the AH of the spacetime metric components in higher-dimensions.


As pointed out above, the fact that the AH that forms in the collisions that we study in this article (see Figs.~\ref{fig:Bin7d}, \ref{fig:Bin6d-1}, \ref{fig:Bin6d-2}), is a non-star-shaped, highly complex and rapidly changing surface, makes the problem of finding it particularly challenging. One possible way forward is to consider a reference surface with the approximate shape of the AH \cite{Pook-Kolb:2018igu,Pook-Kolb:2019ssg}. However, in our case, the reference surface should capture some of the key features that appear as a consequence of the GL cascade; since the latter develops randomly on different scales, both in time and space, finding a suitable reference surface at each time step seems impractical.  Alternatively, one could use an intrinsic parametetrization of the surface and impose some gauge condition (e.g., harmonic) to fix the reparametrization freedom as in \cite{Figueras:2017zwa}. However, in this reference it was found that the resulting system of equations is not very robust in practice in spite of being manifestly elliptic. Therefore, the problem of parametrizing and finding the AHs that generically appear in higher dimensional BH physics remains open. As we will explain below, a promising avenue would be to use certain contours of the evolved variable $\chi$ as the reference surface. This approach has the advantage that these contours capture some of the key features of the AH at any given instant of time and they are readily available since $\chi$ is one of the evolved variables. Implementing this in practice is beyond the scope of this article and we leave it for future work.\footnote{Note that parametrizing a $\chi=\text{const}.$ hypersurface in an efficient and practical way remains challenging because these surfaces are not star-shaped in general.} 

\begin{figure}
\center
\includegraphics[width=0.98\columnwidth]{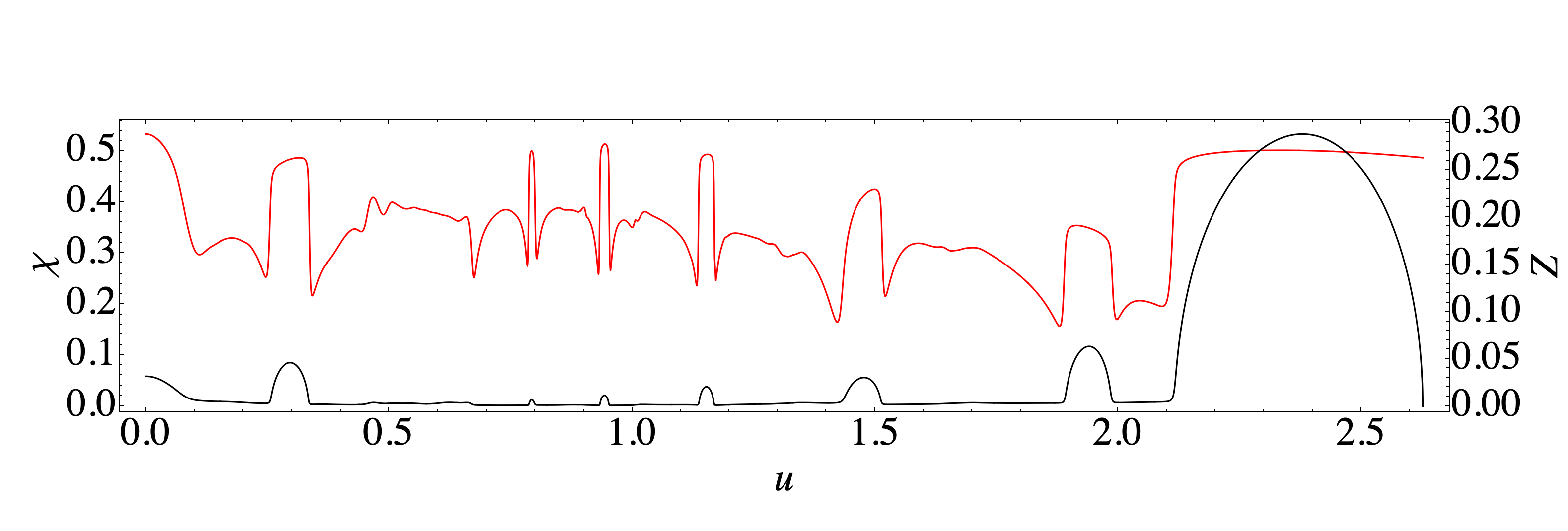}
\caption{%
$\chi$ (in red) on the AH plotted against the embedding coordinate $u$ near the endpoint of the ultraspinning instability of a MP BH in 6$D$ with an initial spin parameter $a/\mu^{1/3}=1.85$ (see \cite{Figueras:2017zwa} for more details); the $Z$ coordinate of the embedding is shown in black and it serves to guide the eye. This figure shows that on the AH,  $\chi\sim 0.5$ on the blobs and $\chi\sim0.3$ on the necks}
\label{fig:chi_AH_1}
\end{figure} 

\begin{figure}
\center
\includegraphics[width=0.98\columnwidth]{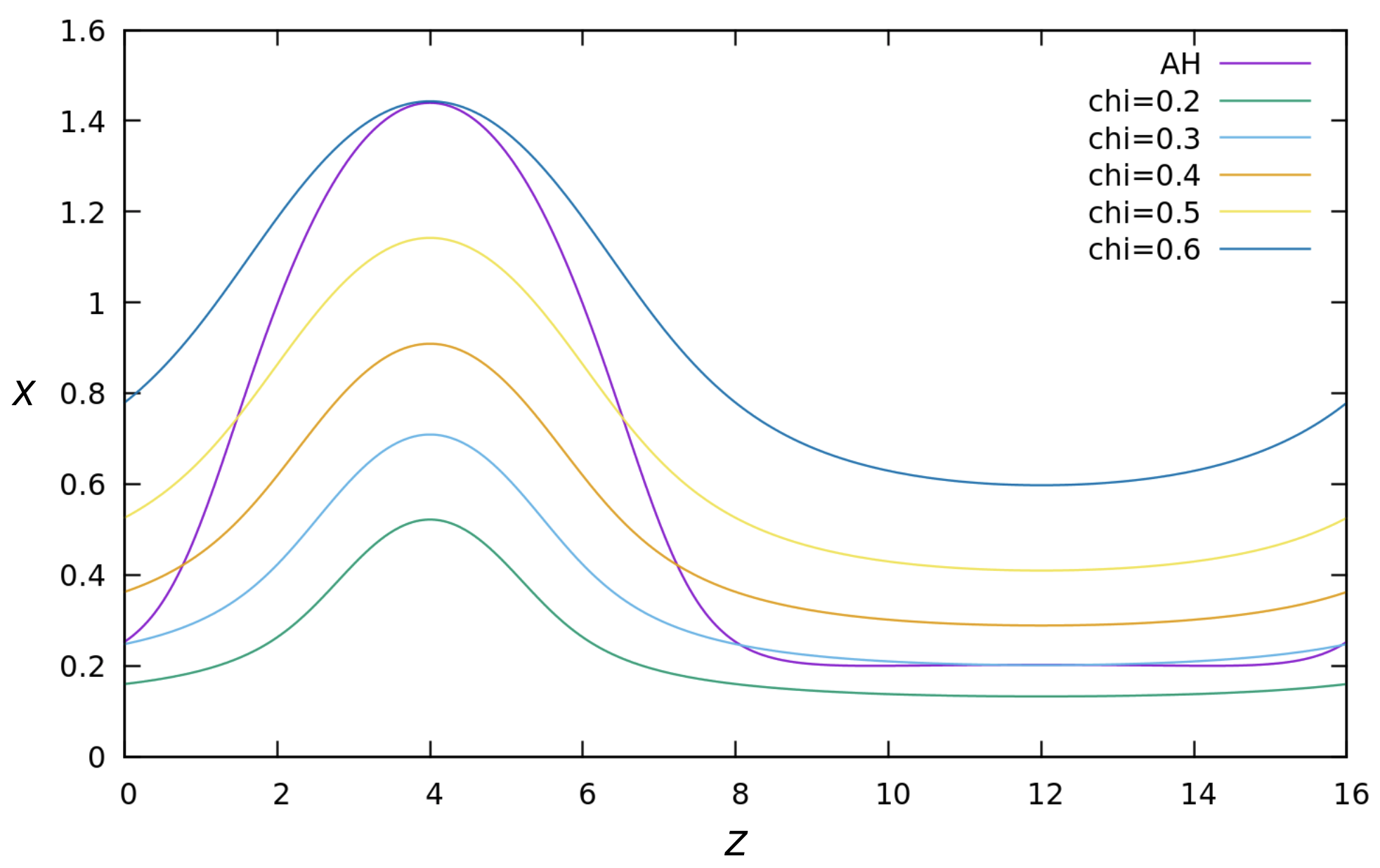}
\caption{%
AH and several $\chi$ contours during the evolution of the GL instability of a black string in 5$D$.  Here $X$ and $Z$ are the coordinates of the computational domain for a black string with a four-dimensional mass parameter $M=1$ and a Kaluza-Klein circle with asymptotic length $L=16$.}
\label{fig:chi_AH_2}
\end{figure}

In Fig. \ref{fig:chi_AH_1} we show the values of $\chi$ on the AH near the endpoint of the ultraspinning instability of a MP BH in 6$D$ with an initial spin parameter $a/\mu^{1/3}=1.85$ reported by one of us in \cite{Figueras:2017zwa}. The shape of the AH is shown in black. This plot shows that the $\chi=0.5$ contours capture the `bulges' quite accurately, especially those from the early generations that have reached a quasi-stationary state. On the other hand, the $\chi=0.3$ contours capture the neck regions of the AH. 
In Fig. \ref{fig:chi_AH_2}  we depict the AH together with different $\chi$ contours at some instant of time during the evolution of the GL instability of a black string in 5D.\footnote{We would like to thank Tiago Fran\c{c}a and Chenxia Gu for kindly sharing their data with us.} This plot shows that indeed these contours of $\chi$ capture the key features of the AH in the evolution of the GL instability, in this case of a black string. 

To get some further intuition about which contours accurately describe different portions of the AH during the GL cascade in a BH collision in $D\geq 6$, we evaluate the normalized Kretschmann invariant on a certain $\chi$ contour and compare its value to that on the horizon of a stationary black string and a spherical black hole. More precisely, we use the definition 
\begin{equation}
\label{norm K}
	\tilde K = \frac{1}{240}\,R^{\alpha \beta \mu \nu} R_{\alpha \beta \mu \nu}\,W(\chi_0)^4
\end{equation}
where $W$ is the proper width of a given $\chi=\chi_0$ contour.
%
The normalization in \eqref{norm K} has been chosen so that in 7$D$, $\tilde K=1$ and $\frac{5}{2}$, respectively, on the horizons of stationary black string and a Schwarzschild black hole. 

Fig. \ref{fig:chi_kret} shows that for the regions of the geometry that have reached a quasi-stationary state, $\tilde K\sim1$ on a $\chi=0.6$ contour and $\tilde K\sim \frac{5}{2}$ on a $\chi=0.7$ contour.  This observation tells us two things: 1) The neck and bulge regions of the AH that have reached a quasi-stationary state are very well approximated by static black strings and spherical black holes respectively, 2) The  constant $\chi$ contours accurately capture the AH on these regions. 
Following these observatiosn, it seems justified to use the constant $\chi$ contours to track the thinning of the neck in the string-like regions and the contour to estimate the masses and angular momenta of the large bulges when we compute the total energy radiated in the collision and subsequent evolution of the system, as we describe in Appendix \ref{app:total_rad}.

We should emphasise  that this argument of comparing Eq.~\eqref{norm K} to the value of the Kretschmann scalar on the horizon of a black string or a spherical black hole is only valid if  the portion of the horizon that one is considering has reached a quasi-stationary state; on the other hand, there is no reason why the constant $\chi$ contours should track the AH in the highly dynamical regions. 

\section{Results}%
\label{sec:results}
For both $D = 6, 7$ we have explored the space of parameters $a$, $v$, $y_0$ by running 
some low resolution simulations until we observe the formation of a long bar-like horizon 
after the merger. As mentioned in the Introduction, no fine tuning is required in order to achieve such states. 
Out of the parameters explored, we have selected three study cases to continue the evolution further with the
aid of higher resolution simulations. 
%
Henceforth we measure time and distance in dimensionless
form using the mass parameter $\mu$ of a single member of our BH binary.
Specifically, we display time
in the form of $t=\tilde{t}/\mu^{1/(D-3)}$, where $\tilde{t}$ is coordinate
time in geometric units, and likewise for radius $r$ and the Cartesian coordinates
$x,\,y,\,z$.

\subsection{$D = 7$}

The $D=7$ collision that we have already discussed in the Introduction (Sec.~\ref{sec:Intro}), is characterized by initial data with parameters $v= 0.5$, $a = 0.7$, $y_0 = 1.1$, $x_0 = 10$.
Note that the spin is within the stability bound $a_{\rm max} = 0.74$ reported in \cite{Dias:2014eua}.
First, we have performed a simulation in a small domain of cubic size $L = 64$, including up to 10 levels. In a second stage, to extract the gravitational waves emitted, we repeated the simulation on a domain with size $L=256$ while keeping the same resolution in the inner levels as in the previous run. 

\subsubsection{Geometry of the AH}
\label{sec:geomAH}

The time evolution obtained in our 7D simulation is displayed in Fig.~\ref{fig:Bin7d} above, which clearly shows that a dumbbell-shaped horizon forms after merger, consisting of two spherical blobs joined by a thin neck.
Moreover, we observe the appearance of spatial modulation along the horizon, characteristic of the GL 
instability. At the same time,  the length of the dumbbell increases making the middle neck thinner, which accelerates the
growth of the instability.  
Continuing the evolution further, we see the formation of two generations of satellites, e.g., small spherical 
beads joined by thin tubes, closely resembling the late stages of unstable strings and ultra-spinning BHs. Our numerics have allowed us to observe the onset of a third generation, in which all string segments present in the second generation, develop their own local GL instability. Note that in 7$D$, the critical value for the width over length ratio of strings which are GL unstable is $r_{{\rm GL}, 7D} \equiv \rm{widht} /\rm{length} \approx 0.5$. We have explicitly checked that the string segments present in our first and second generation are well below this value. 

\begin{figure}[t!]
\centering
\includegraphics[scale=0.4]{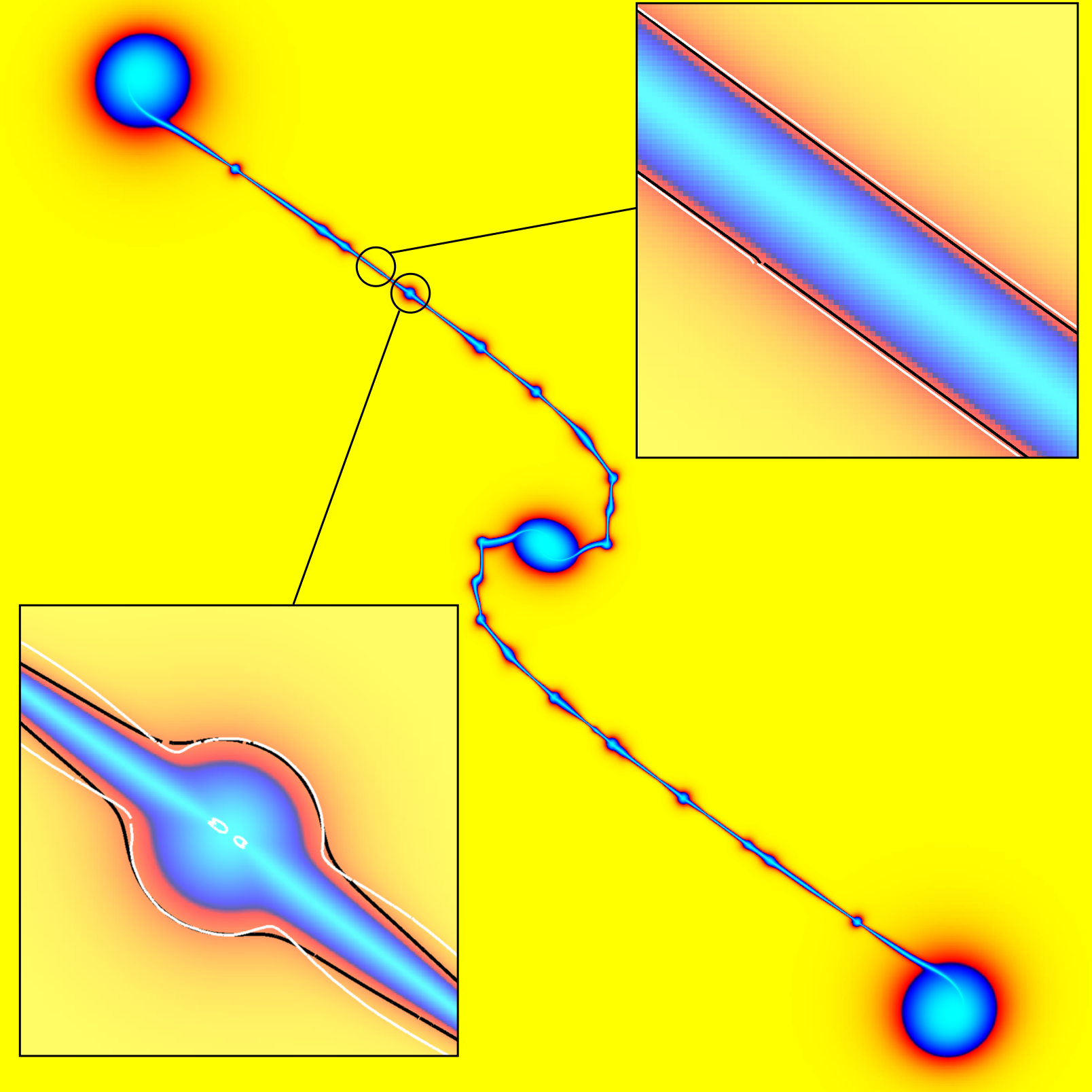}
\caption{Last time slice shown in Fig \ref{fig:Bin7d} for the 7$D$ collision at $\Delta t = 32.75$. We zoom into two different regions of the thinnest part of the geometry. In the string-like region (upper right inset) we show the $\chi = 0.6$ contour as a black line, and the  $\tilde K|_{\chi_0=0.6} = 1$ contour in white. In the region near a satellite (lower left inset) we show the contours $\chi = 0.7$ in black and the  $\tilde K|_{\chi_0=0.7} = 5/2$ contour in white. The fact that these contours are in very good agreement indicates that the contours of $\chi=0.6,0.7$ track the horizons of string-like and spherical-like portions of the geometry respectively. Furthermore, the unnormalized Kretschmann scalar diverges $\propto W^{-4}$, where $W$ is the proper width of the neck-like regions of the geometry. }
\label{fig:chi_kret}
\end{figure}

\begin{figure}
\center
\includegraphics[width=0.98\columnwidth]{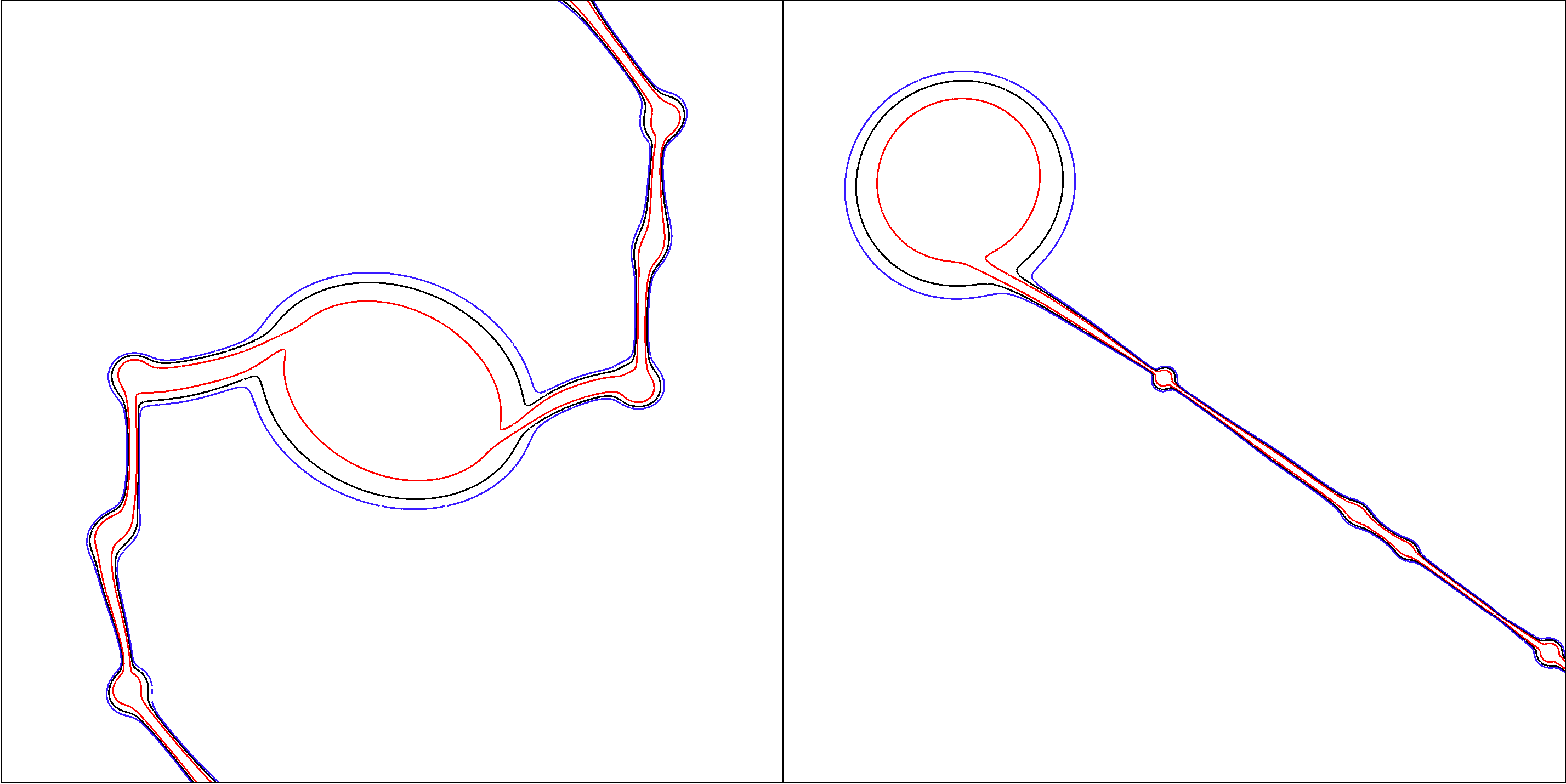}
\caption{%
Comparison of the $\chi = 0.6$ (in blue),  $\chi = 0.5$ (in black) and $\chi=0.3$ (in red) contours for the the 7$D$ BH collision. The $\chi = 0.6$ contour tracks the AH near the neck regions.  The key features of the AH are captured by all these contours; in particular, in the neck regions they are very close to each other. This effect becomes more pronounced the larger $D$ is.}
\label{fig:ChiCont}
\end{figure}

In Fig \ref{fig:chi_kret} we show the last time slice in our 7$D$ simulation, where we overlay contours of the normalized Kretschmann scalar \eqref{norm K} computed at certain $\chi=\chi_0$ surfaces. As the top inset shows, the fact that the contour $\tilde K|_{\chi_0=0.6}=1$ agrees very well with the $\chi=0.6$ contour, indicates that the neck-like regions of the geometry are essentially described by stationary black strings of thickness $W(\chi_0=0.6)$. Furthermore, this implies that the $\chi=0.6$ contours accurately track the horizon on the string-like regions of the geometry.
Finally, the Kretschmann curvature scalar becomes very large inside the $\chi_0=0.6$ contour (see Fig. \ref{fig:K_log} below),
and rapidly grows in time as the neck's proper width $W$ shrinks to zero, with a rate about $\propto W^{-4}$, as also observed for
black strings \cite{Lehner:2010pn,Lehner:2011wc}, and Myers-Perry black holes \cite{Figueras:2017zwa}.  On the other hand, in the lower inset in Fig. \ref{fig:chi_kret}, we see that the $\tilde K|_{\chi_0=0.7} = \frac{5}{2}$ hypersurface agrees very well with the $\chi=0.7$ contour, indicating that the latter tracks the horizon in the bulge regions, which in turn are well-described by quasi-static $7D$ Schwarzschild black holes.  
Moreover, in Fig. \ref{fig:ChiCont} we compare the $\chi = 0.6$ (in blue), $\chi=0.5$ (in black) and the $\chi=0.3$ (in red) contours for the $7D$ BH collision described above. The fact that they all come very close to each other in the string-like region is a consequence of the steep near horizon gradients found in large $D$, and supports the idea that any such constant $\chi$ contour is a good proxy for the AH in the neck regions. 

We have confirmed that for our configurations, the curvature at the AH grows without bound at late times. To do so, in Fig.~\ref{fig:K_log} we display the logarithm of (the absolute value of) the normalised Kretschmann scalar as defined in Eq.~\eqref{norm K} below, choosing as normalization factor the proper thickness of the $\chi=0.6$ contour on the thin neck of the geometry in Fig. \ref{fig:chi_kret}.  The black line in this plot indicates the $\chi=0.6$ contour which, according to the previous discussion, should correspond to the location of the AH in the neck regions of the geometry. This figure clearly shows that  $\tilde K|_{\chi_0=0.6}$ is very large inside the neck regions. Note however that $\tilde K|_{\chi_0=0.6}$ does not diverge; the reason is that in the $(1+\log)$-slicing gauge that we use, the spatial hypersurfaces approach the physical singularity but they actually never touch it. Therefore, this figure shows that our spacelike hypersurfaces behave exactly as one would expect in the moving-puncture gauge.

\begin{figure}
\center
\includegraphics[width=0.6\columnwidth]{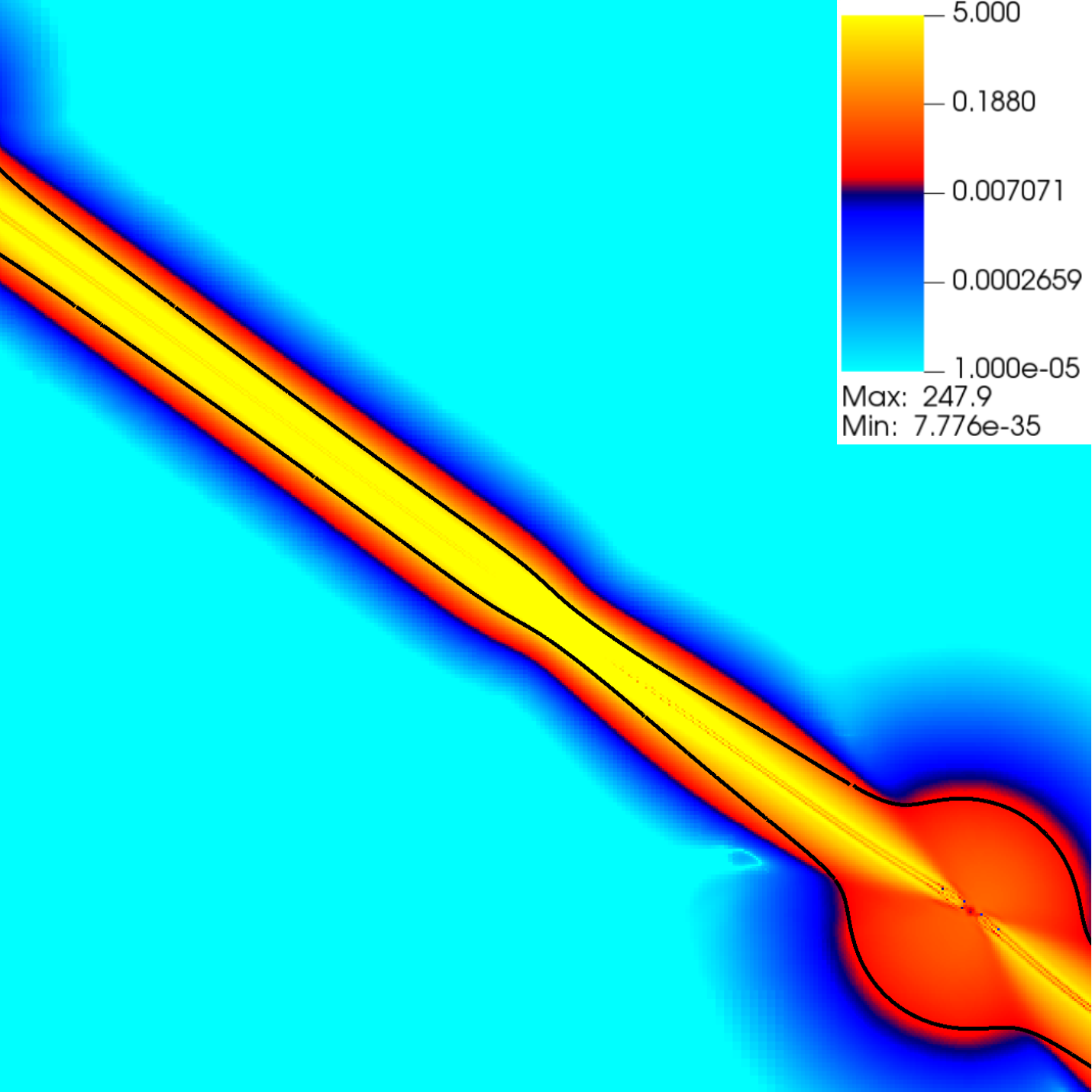}
\caption{%
Logarithm of the absolute value of the normalized Kretschamann scalar $\tilde K|_{\chi_0=0.6}$ (as defined in the main text) for the last time slice of our 7$D$ simulation. Note that this quantity spans several orders of magnitude between the $\chi = 0.6$ contour (black solid line) and the regions far away from it.}
\label{fig:K_log}
\end{figure}

\subsubsection{Gravitational radiation}
In this subsection we consider the gravitational waves emitted during the evolution of the system in $D=7$,   from the initial collision until well into the non-linear evolution of the GL instability. We have relegated the technical details of gravitational wave extraction in 7$D$ to Appendix \ref{app:gw7d}.

\begin{figure}[t!]
\centering
\includegraphics[scale=0.6]{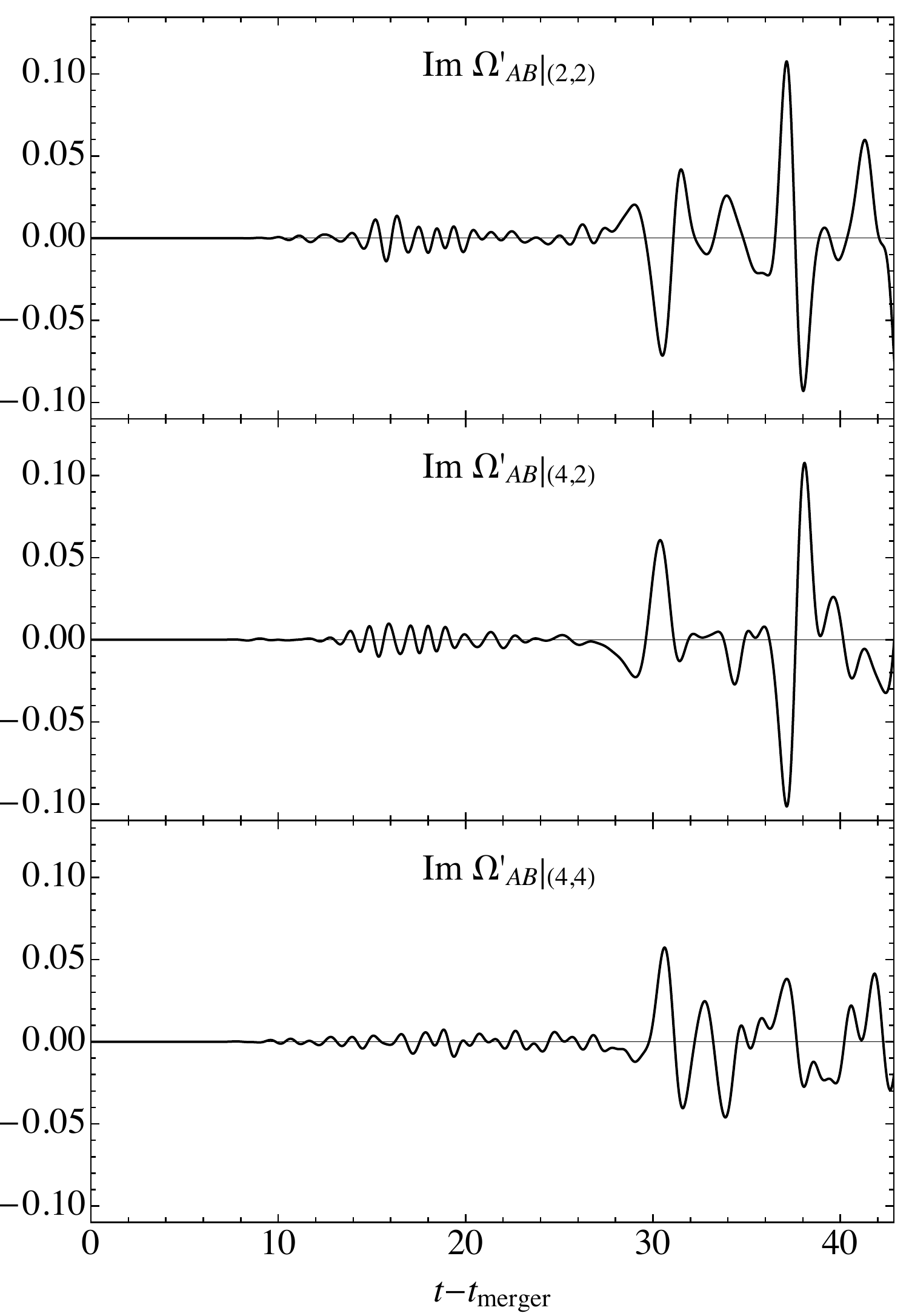}
\caption{Imaginary part of the projection of the matrix of Weyl scalars $\Omega'_{AB}$ onto the harmonics $(\ell_5,m)=(2,2), (4,2), (4,4)$ for the collision in $D=7$.  In this plot, the extraction was done at a radius $R=30$ from the center of the grid. The first peak is observed at $t-t_\text{merger}\sim 30$, corresponding to the initial collision of the two BHs; the oscillations observed before this time are due to the constraint violations in the initial data and initial gauge adjustment. Two large peaks are observed in the $(2,2)$ and $(4,2)$ modes at around $t-t_\text{merger}\sim 37.5$. This corresponds to the time when the dumbbell-shaped horizon stops rotating and starts expanding. Beyond this point,  whilst the gravitational wave signal continues to develop new features corresponding to the onset and further evolution of the local GL instability in the neck regions, their amplitude is never as large as that of these two initial peaks. Consequently, the energy loss in absolute value decreases.}
\label{fig:GWs}
\end{figure}

We display the results for gravitational waves extracted for $D=7$ in Fig. \ref{fig:GWs}.
In this figure, the small oscillations around $t-t_{\rm merger}\lesssim 20$
are due to the initial constraint violations from superposing the two MP BHs and, hence, are discarded as unphysical.
We observe that the first physical peak of radiation in the various modes appears at 
$t-t_\text{merger}\sim 30$, corresponding to the ``collision" of the two BHs, when a common enveloping horizon first forms. 
As shown in the second panel in Fig \ref{fig:Bin7d}, this common horizon is dumbbell shaped. Furthermore, this resulting BH is rotating since it acquires part of the total angular momentum of the initial state.
At some point, which depends on the total initial angular momentum (and the number of dimensions $D$), the arms of the BH stop rotating and the necks connecting the two bulges to central BH start expanding, producing a second radiation peak;
in the $D=7$ example of Fig. \ref{fig:GWs},
this second peak of radiation is visible at around $t -t_\text{merger} = 37.5$. It turns out that this second peak of radiation has the largest amplitude. 
We confirm this picture with a simple model in Appendix \ref{app:gw_toymodel}.

\begin{figure}
\center
\includegraphics[width=0.7\columnwidth]{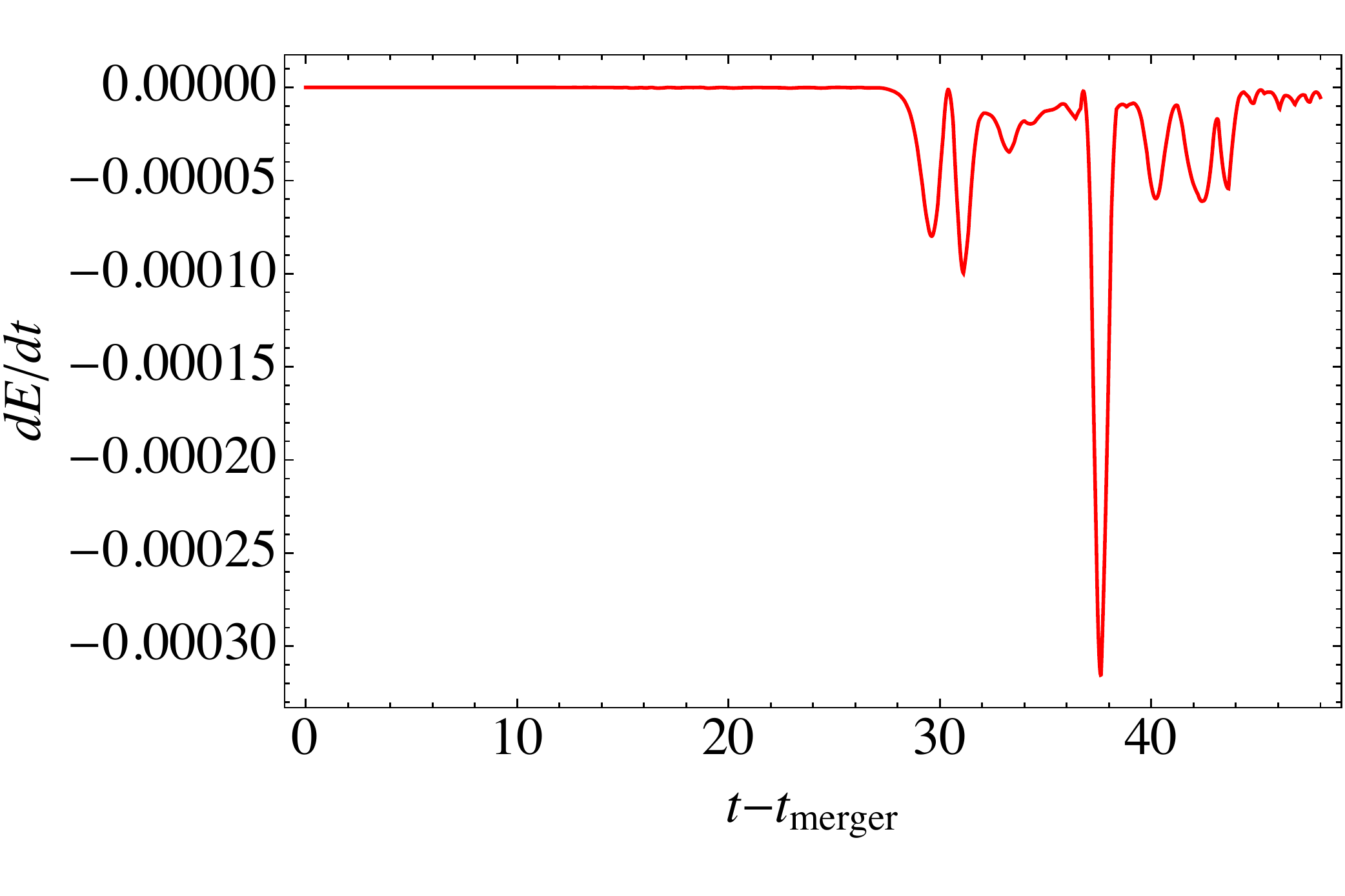}
\caption{%
Energy loss for our $7D$ simulation as computed at a radius $R=30$ from the merger point. The large peak at 
$t-t_\text{merger}\sim 37.5$ is produced when the dumbbell-shaped BH stops rotating and starts expanding.}
\label{fig:Edot}
\end{figure}

We can compute the energy loss $\frac{dE}{dt}$ from the matrix of Weyl scalars according to the formula \cite{Godazgar:2012zq},
\begin{equation}
\label{eq:Edot_Omega}
\frac{dE}{dt} = - \lim_{r\to\infty} \frac{r^{d-2}}{8\pi}\int_{S^{d-2}}\left(\int_{-\infty}^u\Omega'_{AB}(\hat u,r,x)d\hat u\right)^2 d\omega\,.
\end{equation}
where $d\omega$ is the volume element on the $S^{d-2}$.  We display our results for the energy loss in $D=7$ collision  in Fig.~\ref{fig:Edot}. From this figure it follows that up until and including the second peak of radiation, the total radiated energy is about $ 0.01 \%$ of the Arnowitt-Deser-Misner (ADM) mass, which is very small. 
%
We expect that the radiation emitted at later stages to amount to an even smaller fraction, since the dynamics of the outer blobs is quasi-stationary and the GL cascade taking place in the neck regions of the AH produces very little radiation. The reason is that the mass contained in these highly dynamical regions is decreasingly small (Ref.~\cite{Bantilan:2019bvf} reports that less than $5\%$ of the total ADM mass is radiated during the last stage of GL in the 6$D$ evolution of ultraspinning BHs), so while higher modes will be excited, their amplitudes will be decreasing in time. 
We discuss this issue further in Appendix \ref{app:total_rad}, where we provide a crude upper bound on the total energy radiated.  
%

In summary, we have shown that the effects of the gravitational  radiation are sufficiently suppressed in $D \geq7$. In the following subsection we will show that those effects are also suppressed in $D=6$. 
Therefore, we conclude  that there is no mechanism to stop the neck regions to reach zero thickness in finite time. Therefore, the violation of the WCC in BH collisions in $D\geq 6$ is inevitable provided that the initial total angular momentum of the system is sufficiently large.

\subsection{$D = 6$}

In this subsection we discuss the results that we have obtained in $D = 6$. As mentioned above, our results are qualitatively similar to the $D = 7$ case and hence we will be rather brief to avoid repetition.  In summary, in the $D=6$ case  we also see strong evidence for the pinching off of the horizon in finite asymptotic time and hence the violation of the WCC. 

We have run two high resolution simulations in cubic domains of size $L = 64$, with collision parameters
$\{v = 0.5$,  $a = 0.6$, $b = 2.5\}$,
$\{v = 0.45$,  $a = 0.7$,  $b = 2.5\}$,
respectively. Note that both of these values of $a$ are well within the stability regime for rotating MP  BHs $a_{\rm max} = 0.73$ \cite{Dias:2014eua}. In this case we did not carry out the extraction of gravitational waves due to a lack of computational resources, but we expect the waveforms to be qualitatively similar to the $D=7$ case.

\begin{figure}
\centering
\includegraphics[width=0.75\columnwidth]{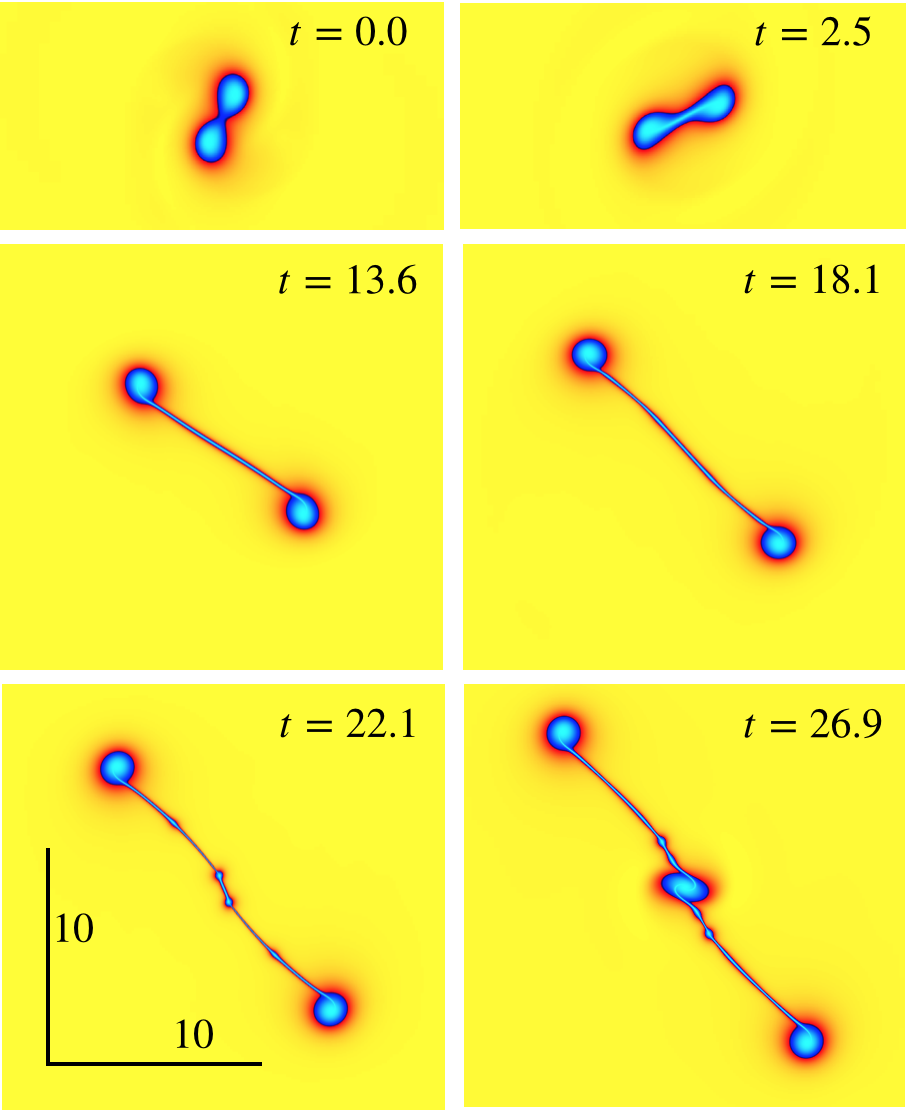}
\caption{%
6$D$-A evolution with $v = 0.5$,  $a = 0.6$, $b = 2.5$.}
\label{fig:Bin6d-1}
\end{figure}

\begin{figure}
\centering
\includegraphics[width=0.55\columnwidth]{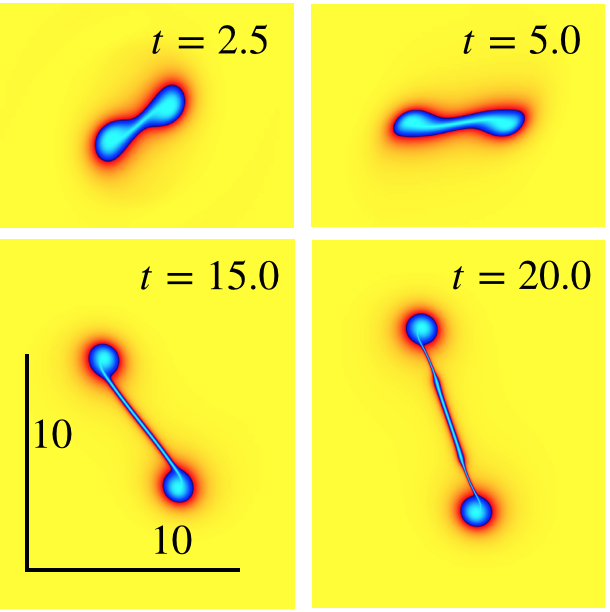}
\caption{%
6D-B evolution with $v = 0.45$,  $a = 0.7$,  $b = 2.5$.}
\label{fig:Bin6d-2}
\end{figure}

In Figs. \ref{fig:Bin6d-1}, \ref{fig:Bin6d-2} we show representative snapshots of the conformal factor $\chi$ on the $z = 0$ plane. 
In the simulation shown in Fig. \ref{fig:Bin6d-1}, the formation of the first generation of satellites is clear, and we observe that the dynamics closely resemble the 7$D$ case. The critical dimensionless ratio for the GL instability in this dimension is $r_{{\rm GL}, 6D} \equiv \rm{width} /\rm{length} \approx 0.4$. Estimating this value for $t = 13.6$ in Fig. \ref{fig:Bin6d-1}, we find that $r_{\rm GL} \approx 0.02$, showing that this string is well in the unstable regime. 
In the dynamics that follow, we again observe the appearance of the central rotating blob, joined to the outer blobs by thin strings. 
Our computational resources did not allow us to reach the same stage for the simulation in Fig.~\ref{fig:Bin6d-2}. However, we do observe a rapid thinning of the neck regions, which should lead to the development of the GL instability as in the previous cases. In particular, for $t = 15.0$ in Fig.~\ref{fig:Bin6d-2} we estimate for $r_{\rm GL} \approx 0.04$, so the strings are unstable. 

As emphasised in the main text, no fine tuning is required to find configurations such as these in which the neck thins and the GL instability is triggered. The observed effects are therefore generic and not due any fine tuning.

\section{Discussion}%
\label{sec:discussion}
By means of numerical GR simulations, we have been able to provide evidence for the generic violation of the WCC conjecture in BH collisions  $D = 6, 7$ spacetime dimensions.
The driving mechanism is the GL instability, and the dynamics unfold 
in close parallel to other known cases of horizon pinching off.  
We have observed the formation of two generations of satellites and the beginning of a third one, which further improves our understanding of the dynamical process. 
We have extracted the gravitational waves emitted by the binaries, showing that
there is a second peak of radiation which surpasses the initial burst due to the merger. 
After this second peak, we expect the emission of radiation to be small,  paving the way for the GL instability 
to break the horizon. 
Since radiation is further suppressed in higher dimensions, our results strongly indicate that 
the violation of WCC we have identified is present in all $D \geq 6$.

It is remarkable that the large $D$ effective theory \cite{Emparan:2015hwa, Emparan:2015gva, Emparan:2016sjk}, 
out of which the arguments for this mechanism were originally drawn \cite{Andrade:2018yqu, Andrade:2019edf}, 
is able to capture the qualitative dynamics so well. 
In hindsight, perhaps the main reason for its success in predicting the violation of WCC 
is the relatively small role that radiation plays already in $6D$. 

As in all other known cases of WCC violation by fragmentation of horizons, the regions which we 
expect to become Planckian are ``small'' at least with respect to the size defined by the geometry of the 
apparent horizon. 
In the language of \cite{Emparan:2020vyf}, they correspond to mild violations of WCC, which means that 
any suitable quantum resolution (analogue of the evaporation of a droplet in hydrodynamics) will have a 
small effect on the subsequent (classical) dynamics. 
In this sense, although the more strict version of the WCC conjecture is violated (naked singularities do appear
in the course of evolution), its spirit is salvaged (the lack of predictability that the violations entail is very small). Note that according to this discussion, the singularity becomes visible because the horizon becomes infinitely thin. As in the previous works \cite{Lehner:2010pn,Lehner:2011wc,Figueras:2015hkb,Figueras:2017zwa,Bantilan:2019bvf}, while with our methods we cannot address the completeness of future null infinity, any horizon penetrating coordinates, such as those of moving-puncture gauge, are suitable to study the approach to the singularity.
More specifically, we show that on the neck-like regions
of the horizon the Kretschmann invariant diverges as the width $W$ of the
neck tends to zero; quantitative investigation furthermore reveals this
blowup to be $\propto W^{-4}$. This the same type of curvature blowup previously observed in black strings \cite{Lehner:2010pn,Lehner:2011wc} and in ultraspinning Myers-Perry black holes \cite{Figueras:2017zwa}, which further supports the fact that underlying mechanism of approach to the singularity is the same in all these systems.
Therefore, in finite time, asymptotic observers will have access to regions of arbitrarily large curvature, thus violating at least in spirit the WCC.

An interesting technical point which we have not addressed here is the derivation of the location of the apparent horizon
from first principles. As discussed above, this requires parametrizing the shape of a reference surface, which is challenging
in scenarios like ours where spatial modulation is constantly changing the shape of the horizon. Obtaining the apparent
horizon in this way would allow us to accurately study entropy production. We leave these interesting issues for future 
research.  

\begin{acknowledgments}
We thank Roberto Emparan for many insightful discussions. Our special thanks are for the entire GRChombo collaboration (www.grchombo.org) for their help and support. TA is supported by the ERC Advanced Grant GravBHs-692951. 
He also thanks QMUL for the hospitality during the early stages of this project. PF is supported by the European Research Council Grant No. ERC-2014-StG 639022-NewNGR, by a Royal Society University Research Fellowship (Grant No. UF140319 and URF\textbackslash R\textbackslash 201026) and by a Royal Society Enhancement Award (Grant No. RGF\textbackslash EA\textbackslash 180260). US acknowledges support from the European Union's H2020 ERC Consolidator Grant ``Matter and strong-field gravity: new frontiers in Einstein's theory'' Grant No. MaGRaTh646597, the STFC Consolidator Grant No. ST/P000673/1, and the GWverse COST Action Grant No. CA16104, ``Black holes, gravitational waves and fundamental physics''.  The simulations presented here were done on the MareNostrum4 cluster at the Barcelona Supercomputing Centre (Grant No. FI-2020-2-0011 and FI-2020-2-0016), the SDSC Comet and TACC Stampede2
clusters through NSF-XSEDE Grant No. PHY-090003, and the Cambridge Service for Data Driven Discovery (CSD3), part of which is operated by the University of Cambridge Research Computing on behalf of the STFC DiRAC HPC Facility (www.dirac.ac.uk). The DiRAC component of CSD3 was funded by BEIS capital funding via STFC capital grants ST/P002307/1 and ST/R002452/1 and STFC operations grant ST/R00689X/1. DiRAC is part of the National e-Infrastructure.

\end{acknowledgments}

\appendix

\section{Initial data}
\label{app:ID}

We begin with a single, un-boosted MP BH with mass and rotation parameters $\mu$, $a$. 
We write the metric as
\begin{equation}
	\label{g KS}
	g_{\mu \nu} = \eta_{\mu \nu} + f(\mathbf{x})\, k_\mu k_\nu\,,
\end{equation}
\noindent where $\eta$ is the Minkowski metric, $f$ is a function of the space-time coordinates $\mathbf{x}$ and $k^\mu$ is a null vector.
For $D$ even, the metric in Kerr-Schild form for a single BH is given by \eqref{g KS} with  
\begin{align}
\label{f1 KS}
	f &= \frac{\mu r}{\Pi(r) F(r, x^i)}\,, \\
	\Pi(r) &= \Pi_{i}^{s(D)} (r^2 + a_i^2)\,,  \\
\label{f2 KS}
	F &= 1 - \sum_{i=1}^{s(D)} \frac{a_i^2 (X_i^2 + Y_i^2)}{(r^2 + a_i^2)} \,,
\end{align}
\begin{align}
	k = dt + \sum_{i=1}^{s(D)} \frac{\{ r( X_i dX_i + Y_i dY_i) + a_i ( X_i dY_i - X_i dY_i) \}}{r^2 + a_i^2}   + \frac{z dz}{r}\,,
\end{align}
\noindent where $k=k_{\mu}dx^{\mu}$, $s(D) = (D-2)/2 $, and $X_i$, $Y_i$
denote subsets of the coordinates that specify the rotation planes; they will be specified explicitly below. The index $i$ runs from $1\ldots (D-2)/2$, and corresponds to the number of possible rotation planes, in such a way that a non-vanishing $a_i$ adds rotation on the axis orthogonal to $(X_i, Y_i)$. We explicitly display the sum over $i$ to emphasise that there is no extra summation of repeated indices. 
The quantity denoted by $r$ is not an independent coordinate, but can be expressed in terms of the $X_i$, $Y_i$ by means of the condition $k_\mu k_\nu \eta^{\mu \nu} = 0$. In order to implement the modified cartoon method in $D = 6$, we choose $a_i = (a, 0)$, $X_i = (x, w_1)$, $Y_i = (y, w_2)$ with $w_1$, $w_2$ being isometric directions. This gives the BH rotation along the $z$-axis. Finally, we use a translation 
$(x',y') = (x-x_0, y-y_0)$ and a Lorentz boost along $x$, 
\begin{equation}
	t'' = t c_\eta + x' s_\eta , \qquad x'' = - t s_\eta + x' c_\eta
\end{equation}
\noindent with $c_\eta = \cosh \eta $, $s_\eta = \sinh \eta $ and $\cosh \eta  = (1 - v^2)^{-1/2}$.

For odd $D$, we have Eqs.~\eqref{f1 KS}-\eqref{f2 KS} along with 
\begin{align}
	k = dt + \sum_{i=1}^{s(D)} \frac{\{ r( X_i dX_i + Y_i dY_i) + a_i ( X_i dY_i - X_i dY_i) \}}{r^2 + a_i^2}  
\end{align}
\noindent and now the number of rotation planes is $s(D) = (D-1)/2 $. 
To implement the cartoon method in $D = 7$, we choose $a_i = (a, 0, 0)$, and  $X_i = (x,z,w_1)$, $Y_i = (y,w_2,w_3)$, 
where $(w_1,w_2,w_3)$ parametrize the isometric directions. This gives the BH rotation along the $z$-axis. 

The mass and angular momentum for a single spinning BH are given by
\begin{align}
\label{MJ MP}
	M = \frac{(D-2) \Omega_{D-2}}{16 \pi G} \mu, \qquad J = \frac{2}{(D-2)} M a  \,,
\end{align}
\noindent where
\begin{equation}
		\Omega_{D-2} = \frac{2 \pi^{(D-1)/2}}{\Gamma[(D-1)/2]}\,.
\end{equation}

\section{Gravitational waves in 7D}
\label{app:gw7d}
%
To extract the gravitational waves in higher dimensions, we have to project the matrix of Weyl scalars $\Omega'_{(A)(B)}$ computed on a certain orthonormal basis onto the basis of spherical harmonics $Y^{(A) (B)}_{\ell \ldots}$ according to \cite{Godazgar:2012zq}:
\begin{equation}
	\Omega'_{\ell \ldots} = \lim_{r \to \infty} r^{(D-2)/2} \int d \Omega_{(n)}  Y^{(A) (B)}_{\ell \ldots} \, ^* \Omega'_{(A) (B)}\,.
\end{equation}
\noindent Here $\ell \ldots$ denotes the set of quantum numbers of the spherical harmonics and $d \Omega_{(n)} $ is the volume element of the $n$-sphere; in our case, in  $7D$ $\ell\ldots = (\ell_5,m)$ and in $6D$,  $\ell\ldots = (\ell_4,m)$. 

As is customary in numerical relativity, in order to carry out the gravitational wave extraction, we have used a finite computational domain of size $L = 256$, 
adding up to eleven levels in order to achieve equal resolutions at the horizons of the BHs as compared to the runs with $L = 64$.
We have extracted the waves at radii $R = 25.0, 27.5, 30.0$ and observed the expected linear behaviour: all three wave forms match in all channels after the corresponding time shifts.

We now describe our basis for the tensor spherical harmonics on the $S^5$. The relevant tensor harmonics on the  $S^4$ needed for the analogous computation of the gravitational waves in the 6$D$ collisions can be found in Ref.~\cite{Bantilan:2019bvf}.

It is convenient to write the metric on  the $S^5$ at infinity such that it has a manifest $SO(4)$ symmetry and one of its $U(1)$'s is aligned with the $U(1)$ on the rotation plane of the incoming MP BHs:
\begin{equation}
ds^2 = d\chi^2 + \sin^2\chi\,d\phi^2 + \cos^2\chi\,d\Omega_{(3)}^2\,,
\label{eq:metS5}
\end{equation}
where $d\Omega_{(3)}^2$ is the metric on the unit $S^3$.\footnote{In this appendix we use $\chi$ to denote the polar angle on the $S^5$; it should not be confused with the conformal factor of the induced metric on the spacelike hypersurfaces.} As shown in Ref.~\cite{Bantilan:2019bvf}, due to the symmetries of the spacetime, the gravitational wave signal is contained in the scalar derived tensor harmonics. Given the form of the metric \eqref{eq:metS5}, a convenient basis for the scalar harmonics on the $S^5$ is
\begin{equation}
\begin{aligned}
\mathbb Y^{\ell_5, m,\ell_3,\ldots}  =&~N\,e^{\text{i}\,m\phi}(\sin\chi)^{|m|}(\cos\chi)^{\ell_3}\mathbb Y^{\ell_3\,\ldots}\\&\times\,_{2}F_1\big(\textstyle{\ell_3+|m|-k,k+\frac{3}{2},\ell_3+\frac{2}{2};\cos^2\chi}\big)\,,
\end{aligned}
\label{eq:scalarYS5}
\end{equation} 
where $_{2}F_1$ is a hypergeometric function,  $N$ is a normalization factor, $k$ is a non-negative integer and $\mathbb Y^{\ell_3\,\ldots}$ are the scalar harmonics on the $S^3$, so that $\Delta_{S^3}\mathbb Y^{\ell_3\,\ldots}=-\ell_3(\ell_3+2)Y^{\ell_3\,\ldots}$; the dots in $\mathbb Y^{\ell_3\,\ldots}$ denote the remaining labels for the quantum numbers of the harmonics. Regularity of the harmonics implies
\begin{equation}
\ell_5 = 2\,k-(\ell_3+|m|)\,.
\end{equation}

Since the spacetimes that we consider in this paper preserve a round $S^3$ inside the $S^5$, to extract the gravitational waves we only need to consider harmonics on the $S^5$ that also preserve a round $S^3$. Therefore, from now on, in \eqref{eq:scalarYS5} we will consider harmonics with $\ell_3=0$ and hence we can uniquely label them with only two quantum numbers, $(\ell_5,m)$.

Scalar derived traceless tensor harmonics on the $n$-sphere can be constructed from the scalar harmonics by\footnote{There is a typo in the analogous formula for the scalar derived tensor harmonics in \cite{Bantilan:2019bvf}, which we correct here.}
\begin{equation}
\begin{aligned}
\mathbb S^{\ell_n,\ldots,\ell_1}_{ab} =&~\frac{\sqrt{n}}{\sqrt{(n-1)(\ell_n-1)(\ell_n+n)}}\\
&\times\left(\nabla_a \mathbb S^{\ell_n,\ldots,\ell_1}_b+\frac{\sqrt{\ell_n(\ell_n+n-1)}}{n}\,g_{ab}\,\mathbb S^{\ell_n,\ldots,\ell_1}\right)\,,
\end{aligned}
\label{eq:tensorYs}
\end{equation}
where $S^{\ell_n,\ldots,\ell_1}_a \equiv \frac{1}{\sqrt{\ell_n(\ell_n+n-1)}}\nabla_a S^{\ell_n,\ldots,\ell_1}$ are the scalar derived vector harmonics, and $g_{ab}$ is the metric on the unit $S^n$. The coefficients are chosen so that the scalar derived tensor harmonics are suitably normalized:
\begin{equation}
\int_{S^n}\,g^{ac}\,g^{bd}\,S^{\ell_n,\ldots,\ell_1}_{ab}\,S^{\ell'_n,\ldots,\ell'_1}_{cd}=\delta_{\ell_n\ell_n'\ldots\ell_1\ell'_1}\,.
\end{equation}

To do the extraction, we need to compute the tensor harmonics  on the $S^5$ in a certain basis of angular vectors. For the metric in \eqref{eq:metS5}, we choose the obvious basis, 
\begin{equation}
m_{(1)}=\frac{\partial}{\partial\chi}\,,\quad m_{(2)}=\frac{1}{\sin\chi}\frac{\partial}{\partial\phi}\,,\quad m_{(i)} =\frac{1}{\cos\chi}\,\hat m_{(i)} \,,
\label{eq:basisS5}
\end{equation}
where $\hat m_{(i)}$, $i=1,2,3$ are a basis of angular (unit) vectors on the $S^3$.

After these preliminaries, we are now ready to list the non-vanishing components of the scalar derived tensor harmonics on the $S^5$ that we have used to extract the gravitational waves:
\begin{itemize}
\item $(\ell_5,m)= (2,0)$:
\begin{equation}
\begin{aligned}
\mathbb S^{(2,0)}_{11} =&~ \frac{1}{\pi ^{3/2}}\sqrt{\frac{3}{70}} \big[2 \cos (2 \chi )+1\big] \,,\\
\mathbb S^{(2,0)}_{22} =&~-\frac{1}{\pi ^{3/2}}\sqrt{\frac{3}{70}} \big[\cos ^2(\chi )-4\big]\,,\\
\mathbb S^{(2,0)}_{ij} =& -\frac{\delta_{ij}}{2 \pi ^{3/2}}\sqrt{\frac{3}{70}} \big[\cos (2 \chi )+3\big]\,.
\end{aligned}
\end{equation}

\item $(\ell_5,m)= (2,2)$:
\begin{equation}
\begin{aligned}
\mathbb S^{(2,2)}_{11} =&~\frac{1}{\sqrt{70}\pi ^{3/2}}\,e^{2 \text{i} \phi } \big[2 \cos (2 \chi )+3\big]\,,\\
\mathbb S^{(2,2)}_{12} =&~\frac{\text{i}}{\pi ^{3/2}}\sqrt{\frac{5}{14}}\, e^{2 \text{i} \phi } \cos (\chi )\,,\\
\mathbb S^{(2,2)}_{22} =&~-\frac{1}{2 \sqrt{70} \pi ^{3/2}}\,e^{2 \text{i} \phi } \big[\cos (2 \chi )+9\big]\,,\\
\mathbb S^{(2,2)}_{ij} =&~\frac{\delta_{ij}}{\sqrt{70} \pi ^{3/2}}\,e^{2 \text{i} \phi } \sin ^2(\chi )\,.
\end{aligned}
\end{equation}

\item $(\ell_5,m)= (4,0)$:
\begin{equation}
\begin{aligned}
\mathbb S^{(4,0)}_{11} =&-\frac{1}{2 \sqrt{10} \pi ^{3/2}}\big[ \cos (2 \chi )+5 \cos (4 \chi )-2\big]\,,\\
\mathbb S^{(4,0)}_{22} =& \frac{1}{8 \sqrt{10} \pi ^{3/2}}\big[ -24 \cos (2 \chi )+5 \cos (4 \chi )+3 \big]\,,\\
\mathbb S^{(4,0)}_{ij} =& \frac{\delta_{ij}}{24 \sqrt{10} \pi ^{3/2}}\big[ 28 \cos (2 \chi )+15 \cos (4 \chi )-11 \big]\,.
\end{aligned}
\end{equation}

\item $(\ell_5,m)= (4,2)$:
\begin{equation}
\begin{aligned}
\mathbb S^{(4,2)}_{11} =&-\frac{1}{4 \sqrt{5} \pi ^{3/2}}\,e^{2 \text{i} \phi } \big[2 \cos (2 \chi )+5 \cos (4 \chi )-2\big]\,,\\
\mathbb S^{(4,2)}_{12} =&\frac{\text{i} \sqrt{5}}{16 \pi ^{3/2}}\,e^{2 \text{i} \phi } \big[\cos (\chi )-5 \cos (3 \chi )\big]\,,\\
\mathbb S^{(4,2)}_{22} =&\frac{1}{16 \sqrt{5} \pi ^{3/2}}\,e^{2 \text{i} \phi } \big[2 \cos (2 \chi )+5 \cos (4 \chi )+13\big]\,,\\
\mathbb S^{(4,2)}_{ij} =& -\frac{\delta_{ij}}{4 \sqrt{5} \pi ^{3/2}} \,e^{2 \text{i} \phi }\, \sin ^2(\chi ) \big[5 \cos (2 \chi )+6\big]
\end{aligned}
\end{equation}

\item $(\ell_5,m)= (4,4)$:
\begin{equation}
\begin{aligned}
\mathbb S^{(4,4)}_{11} =&\frac{1}{2 \sqrt{2} \pi ^{3/2}}\,e^{4 \text{i} \phi } \sin ^2(\chi ) \big[2 \cos (2 \chi )+3\big]\,,\\
\mathbb S^{(4,4)}_{12} =&\frac{5 \, \text{i}}{2 \sqrt{2} \pi ^{3/2}}\,e^{4 \text{i} \phi } \sin ^2(\chi ) \cos (\chi )\,,\\
\mathbb S^{(4,4)}_{22} =&-\frac{1}{4 \sqrt{2} \pi ^{3/2}}\,e^{4 \text{i} \phi } \sin ^2(\chi ) \big[\cos (2 \chi )+9\big]\,,\\
\mathbb S^{(4,4)}_{ij} =&\frac{1}{2 \sqrt{2} \pi ^{3/2}}\,e^{4 \text{i} \phi } \sin ^4(\chi )\,.
\end{aligned}
\end{equation}

\end{itemize}
The harmonics with negative $m$ can be obtained from the ones listed above by complex conjugation.  

Finally, to do the extraction we compute the components of the matrix of Weyl scalars $\Omega'_{(A)(B)}$ in a certain orthonormal basis. We construct the latter starting from
\begin{equation}
\begin{aligned}
m_{(1)} =&~ x\,\partial_x + y \,\partial_y + z\,\partial_z\,,\\
m_{(2)} =&~ x\,z\,\partial_x + y\,z\,\partial_y  -(x^2+y^2)\,\partial_z\,,\\
m_{(3)} =& -y\,\partial_x + x\,\partial_y\,, 
\end{aligned}
\label{eq:basisWeyl}
\end{equation} 
and carrying out a standard Gram-Schmidt orthonormalization.\footnote{Note that this basis differs from the one used in \cite{Bantilan:2019bvf}. The reason is explained below.} The unit vectors on the transverse sphere direction are simply given by
\begin{equation}
m_{(\hat i)} = \frac{1}{\sqrt{\gamma_{ww}}}\,\partial_{w_i}\,.
\end{equation}
One can show that, asymptotically, the angular basis vectors obtained from \eqref{eq:basisWeyl} are aligned with \eqref{eq:basisS5}. 



\section{Radiation peak due to separation}
\label{app:gw_toymodel}

In order to elucidate the origin of the second peak that we observe in the pattern on gravitational radiation, 
we consider a simple model for a higher dimensional binary and use the quadrupole formula of \cite{Cardoso:2002pa}. 
Since a general expression is only available for even dimensions, we consider the case $D = 6$ which we 
expect will best capture the qualitative features of our setting.  

%
The corresponding quadrupole formula for even $D$ can be written as \cite{Cardoso:2002pa}
\begin{align}\label{eq:radPower}
\frac { d E } { d t } = P(D) \left[ ( D - 1 ) \partial _ { t } ^ { \frac{ D +2 }{2} } M _ { i j } ( t ) \partial _ { t } ^ { \frac{ D +2 }{2}} M _ { i j } ( t ) - \left| \partial _ { t } ^ { \frac{ D +2 }{2} } M _ { i i } ( t ) \right| ^ { 2 } \right]\,.
\end{align}
\noindent where the prefactor is given by
\begin{equation}
	P(D) = \frac{2^{2-D} (D-3) D}{\pi^{\frac{D-5}{2}} \Gamma\left[\frac{D-1}{2}\right]\left(D^{2}-1\right)(D-2)} G\,,
\end{equation}
and 
\begin{equation}
	M^{ij} =\int  d^{D-1}x\, T^{00}(t,x) x^i x^j\,.
\end{equation}
In addition, \cite{Andrade:2019edf} provided an expression for the emission of angular momentum in even dimensions, which in our case reduces to
\begin{align}\label{eq:Jdot}
\frac { d J^{ij} } { d t } = \frac{4(D^2-1)}{D+2}P(D) \left[ \partial _ { t } ^ { \frac{ D +2 }{2} } Q^{[i} \, _{k} \partial_t ^{\frac{D}{2} } Q^{j]} \, _k  \right]\,,
\end{align}
\noindent where 
\begin{equation}
	Q^{ij} = M^{ij} - \frac{1}{D-1}\delta_{ij} M_{kk}\,.
\end{equation}

\begin{figure}[t!]
\center
\includegraphics[width=0.6\columnwidth]{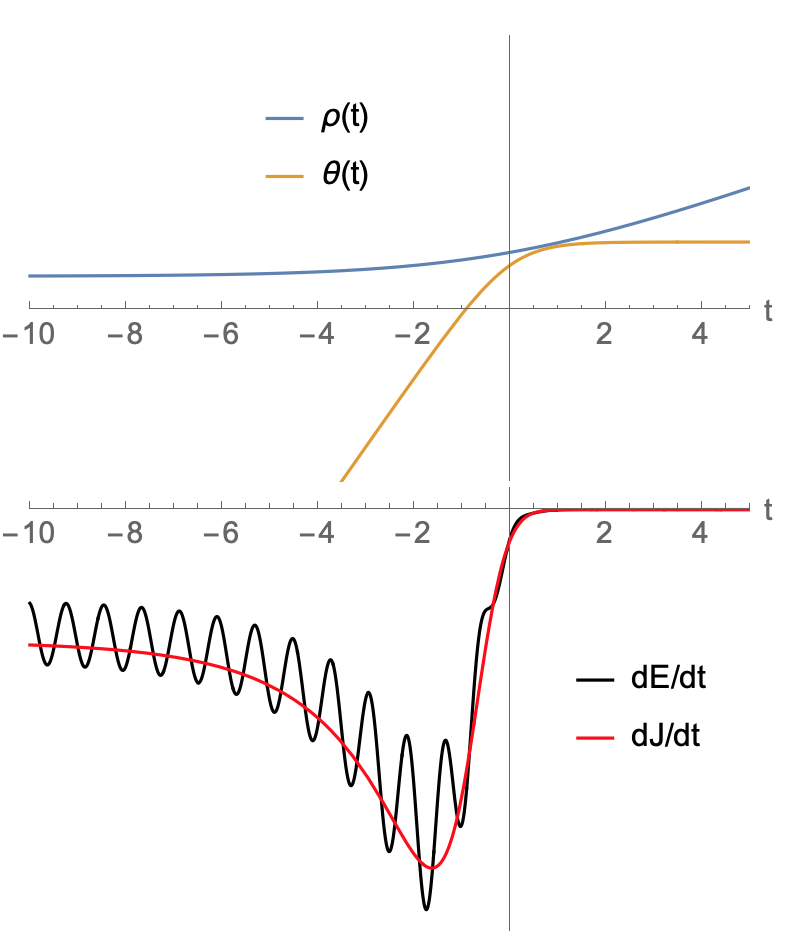}
\caption{%
\textit{Top}: Trajectory for a binary which starts in a circular orbit and experiences a slow down and separation. (Bottom): Energy and angular momentum loss as computed from the quadrupole formulae \eqref{eq:radPower}, \eqref{eq:Jdot} in $6D$.}
\label{fig:gw_toy}
\end{figure}

In order to evaluate these expressions, we make the assumption that the energy distribution $T^{00}$ is simply a delta function with support on two point particles that follow a trajectory on a two-dimensional plane. We parametrize the trajectory in polar coordinates $x_1(t) = \rho(t) \cos \theta(t), x_2(t) = \rho(t) \sin \theta(t) $.
For concreteness, we consider a circular binary which slows down and separates along a straight line. This captures some of the qualitative features we observe in our simulations after merger. We show the profiles we have chosen for $\rho$ and $\theta$ in Fig \ref{fig:gw_toy} (top) while the obtained pattern for energy and angular momentum loss is depicted on Fig \ref{fig:gw_toy} (bottom). 
At early times we observe a sinusoidal pattern for energy radiation characteristic of a circular binary. Interestingly, the later slow down and separation leads to a relatively sharp peak, which resembles the second radiation peak we observe in Fig \ref{fig:Edot}. We also observe that the radiated angular momentum follows the pattern of the  radiated energy averaged over the orbits.

\section{Radiated energy and angular momentum}
\label{app:total_rad}
%
We approximate the total initial and final mass and angular momentum
according to
\begin{align}
	M_i &= 2 \gamma_i m_i \,,\\
	M_f &= 2  \gamma_f m_f + m_{central} + 2 m_{string} \,,\\
	J_i &= \gamma_i m_i \left[ b_i v_i + 2 \frac{2}{D-2} a_i  \right] \,,\\ 
	J_f &=  \left[ \gamma_f  m_f b_f v_f + \frac{2}{D-2} m_{central} a_{central}  \right]\,.
\end{align}
Here $m_i$ is the mass of the BHs in the initial state, with velocities $v_i$, spins $a_i$ and impact parameter $b_i$ -- defined as 
the distance between the center of the BHs and the plane along which the binary is reflection symmetric, i.e.~$b_i = y_0$ in the notation of the main text --, $m_f$ is the mass of the BHs at the extremes of the final dumbbell, which have velocities $v_f$ and impact parameter $b_f$ (we have checked that they do not have intrinsic spin), $m_{central}$, $a_{central}$ are the mass and spin of the central BH, and $m_{string}$ is the mass of each of the strings joining the central and outer BHs. 

We assume the BHs and strings involved in this calculation to be stationary, in which case we can compute their properties from Eq.~\eqref{MJ MP} and (we set $G = 1$ henceforth)
\begin{align}
	M_{\rm string} &= \frac{(D-2) \Omega_{D-2}}{16 \pi G} r_+ L\,.
\end{align}
The initial parameters are given by (recall $\mu = 1$), 
\begin{equation}
	m_i = 3.1, \qquad v_i = 0.5, \qquad b_i = 2.2, \qquad a_i = 0.7\,.
\end{equation}
For the final parameters, we can measure $v_f$ and $b_f$ by tracking the position of the outgoing blobs, and extracting the metric components in the region of the corresponding blobs. We find
\begin{equation}
	v_f \approx 0.3, \qquad b_f \approx 4.26\,.
\end{equation}
We see that the outgoing BHs do not rotate, so they can be approximated Schwarzschild BHs. Their mass is then given by
\begin{equation}
	m_f \approx 3.1\,.
\end{equation}
The mass and rotation parameter of the central blob can be estimated by computing $\Omega = \beta^{\phi} \approx 1.5$, $g_{\phi \phi} \approx 0.25$ and comparing this with the values for a MP BH. We obtain 
\begin{equation}
	m_{central} \approx 0.07, \qquad  J \approx 0.009, \qquad a_{central} \approx 0.32\,.
\end{equation}
Collecting these results, we obtain 
\begin{equation}
	M_i \approx 7.1,  \, J_i \approx 5.9, \qquad  M_f \approx 6.6,  \, J_f \approx 4.2\,.
\end{equation}
Assuming that there could be an uncertainty of $10 \%$ in the values of $m_f$, $v_f$ and $b_f$, we obtain that 
about $8^{+10}_{-10} \%$ of the total mass-energy and $30^{+19}_{-25} \%$ of the angular momentum are radiated. 

\section{Numerical tests}
\label{app:numtest}

%
\begin{figure*}
  \includegraphics[width=0.45\columnwidth]{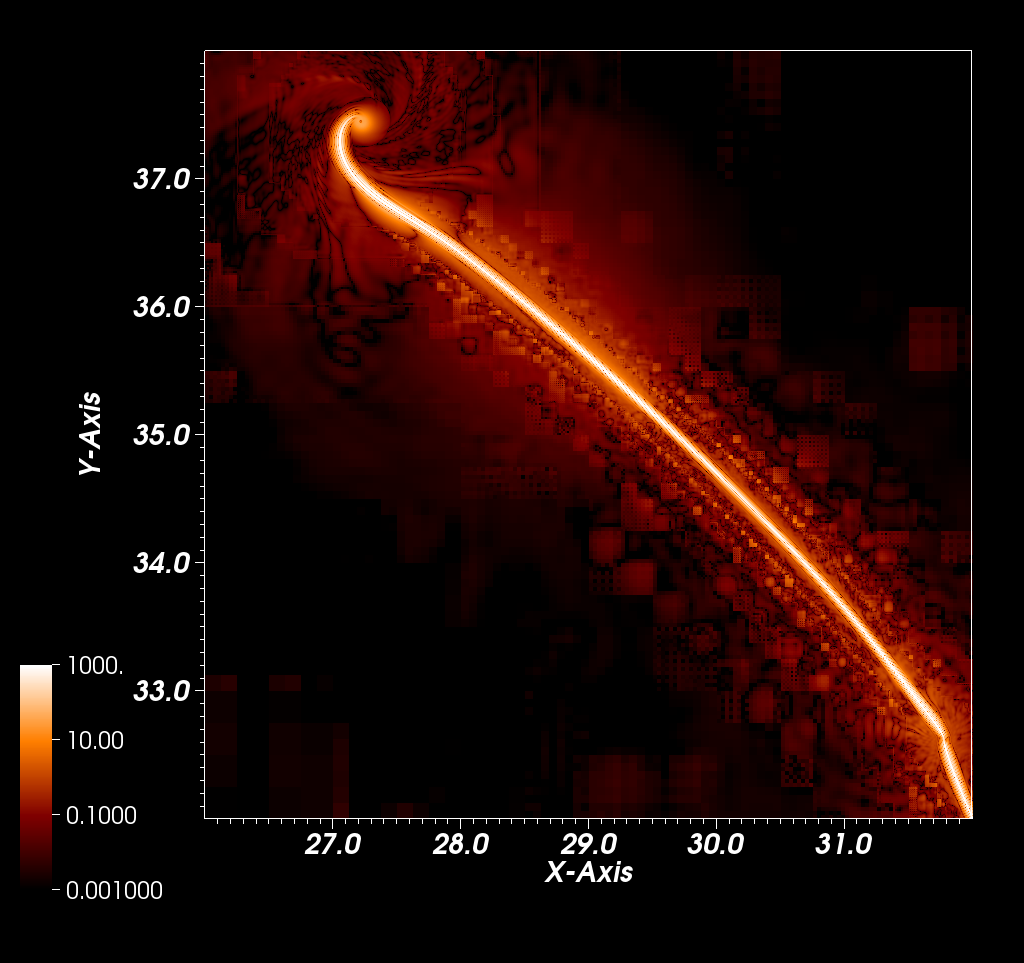}
  \hspace{30pt}
  \includegraphics[width=0.45\columnwidth]{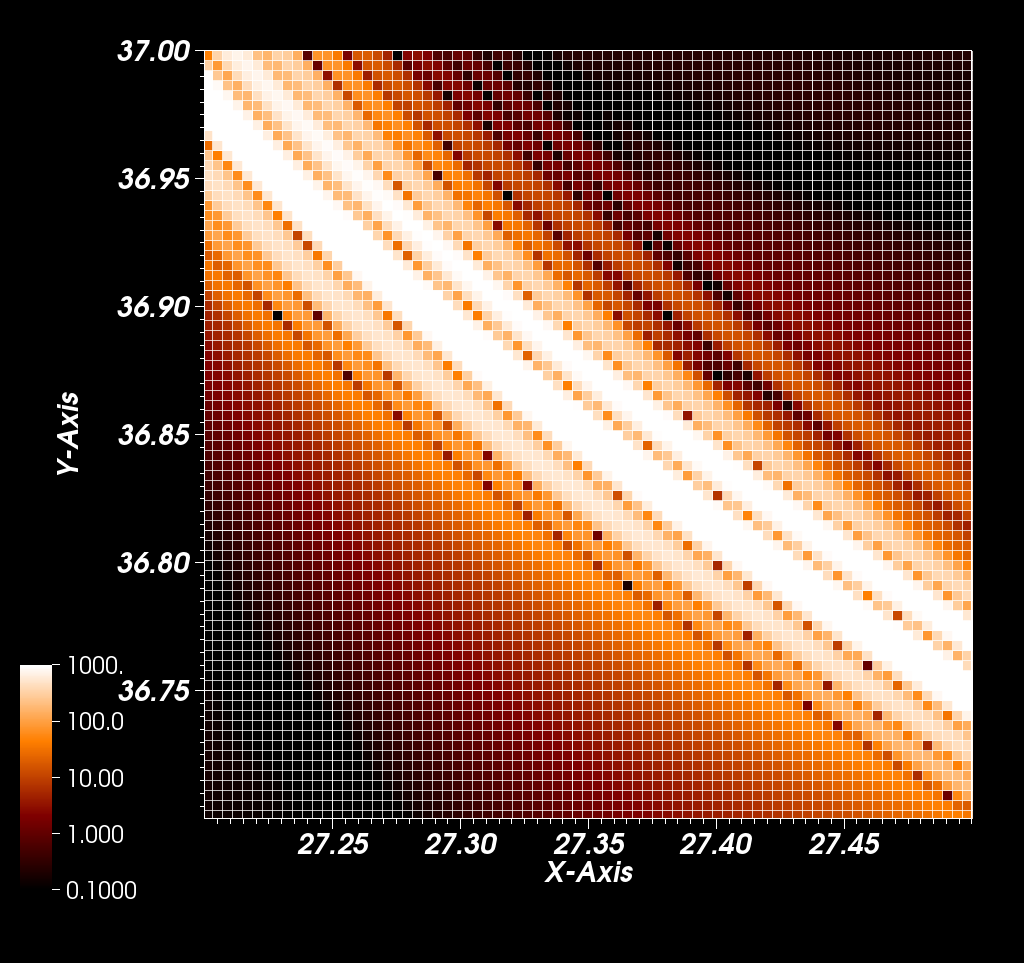}
  \includegraphics[width=0.45\columnwidth]{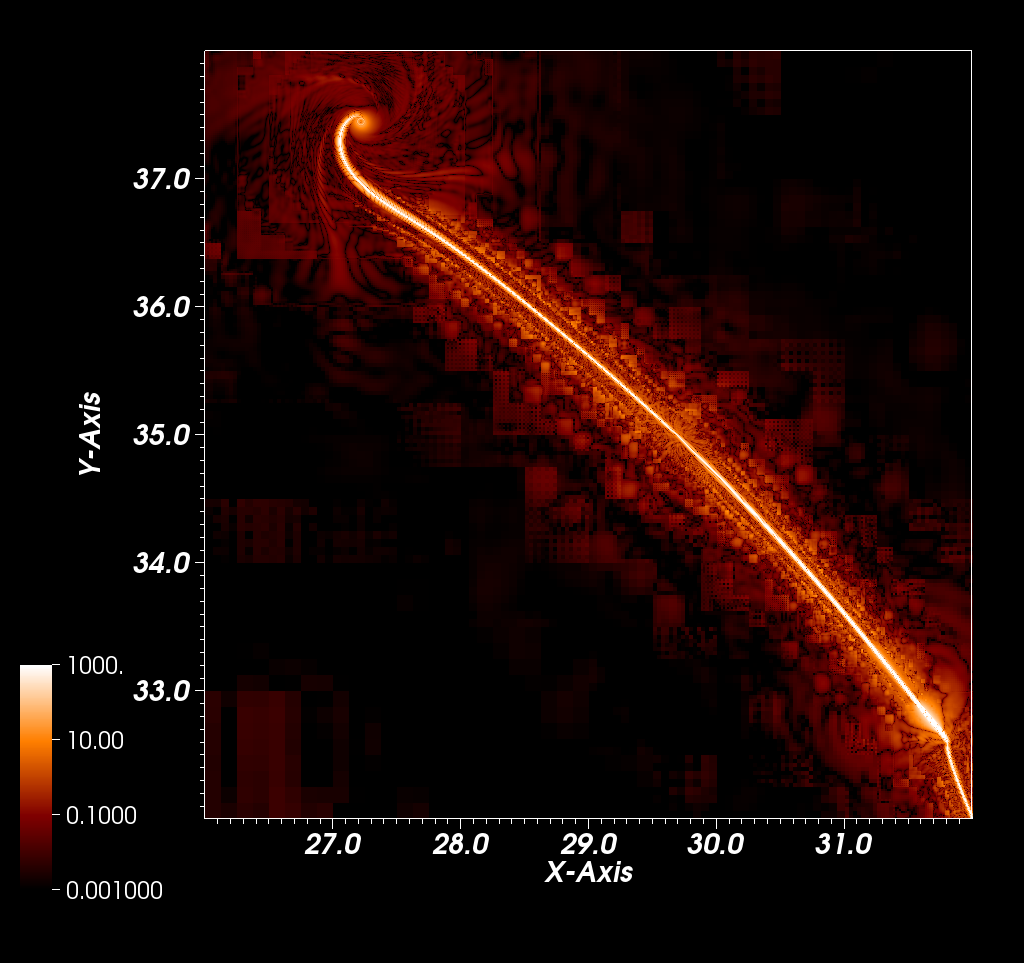}
  \hspace{30pt}
  \includegraphics[width=0.45\columnwidth]{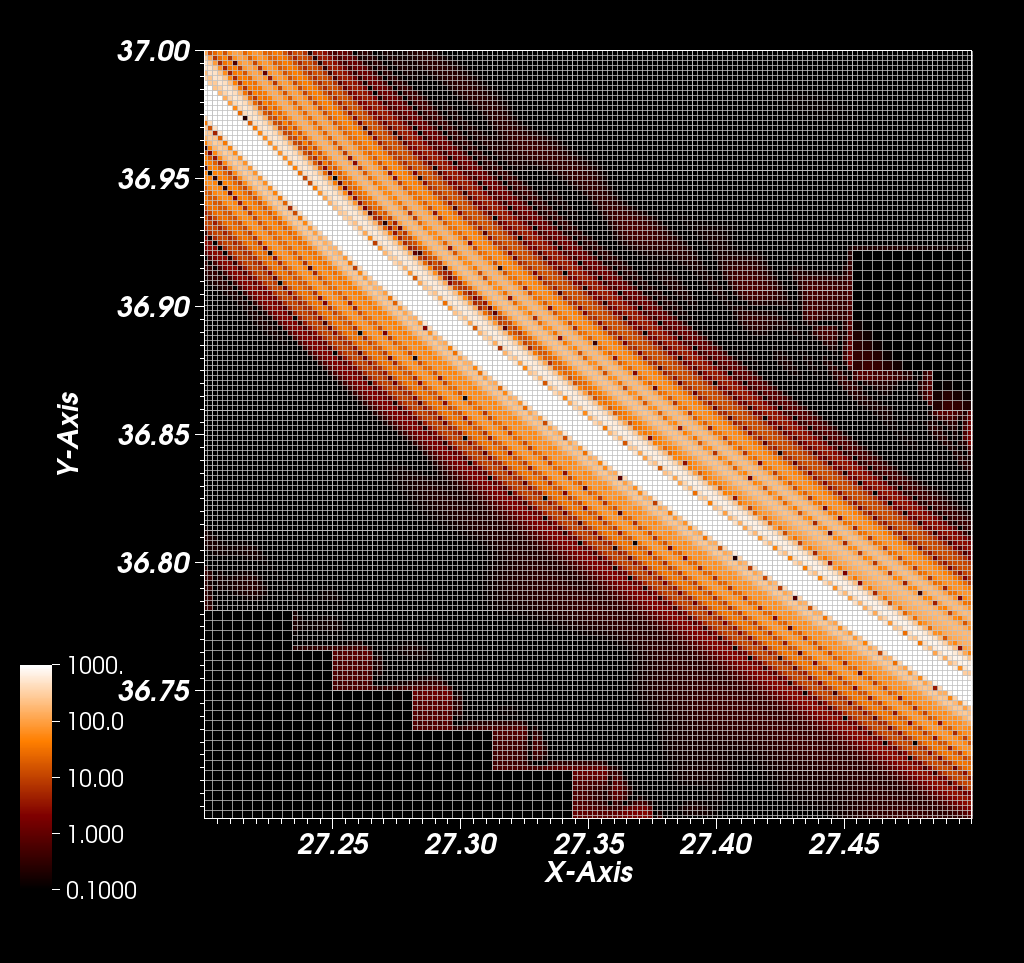}
  \includegraphics[width=0.45\columnwidth]{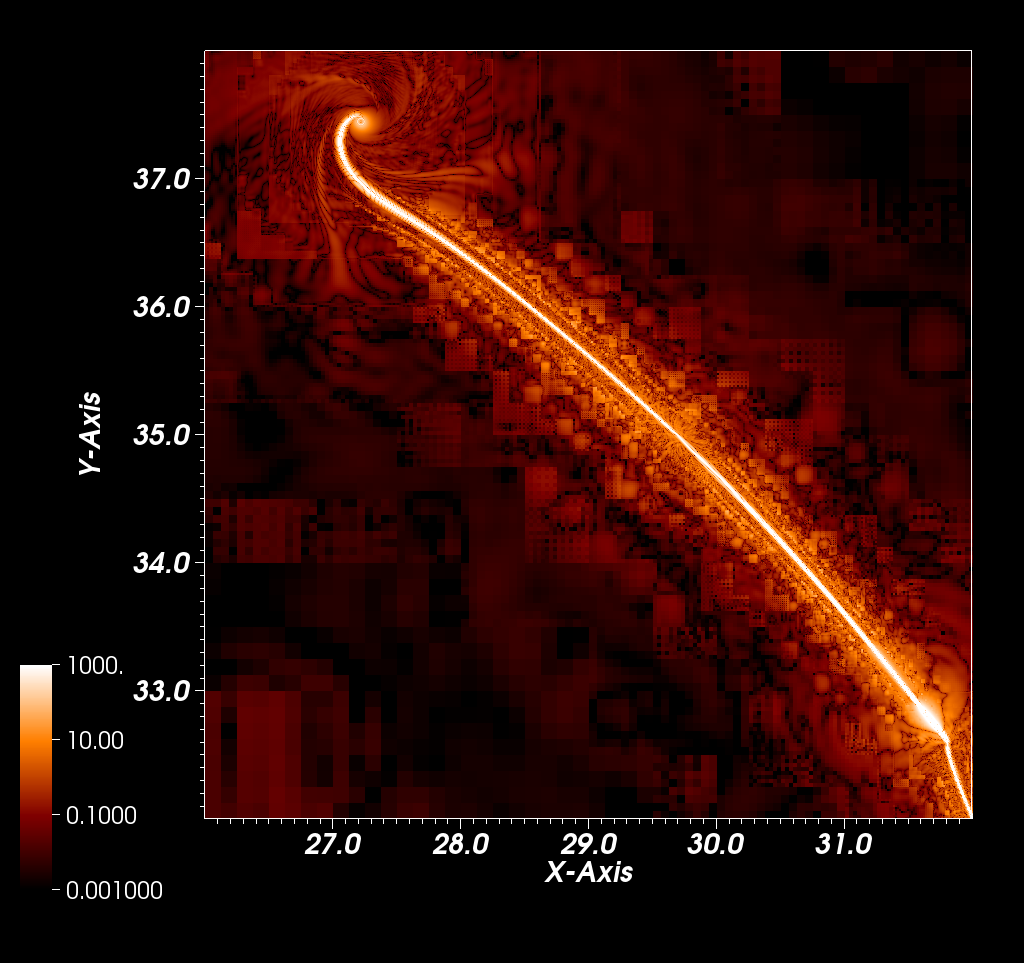}
  \hspace{36pt}
  \includegraphics[width=0.45\columnwidth]{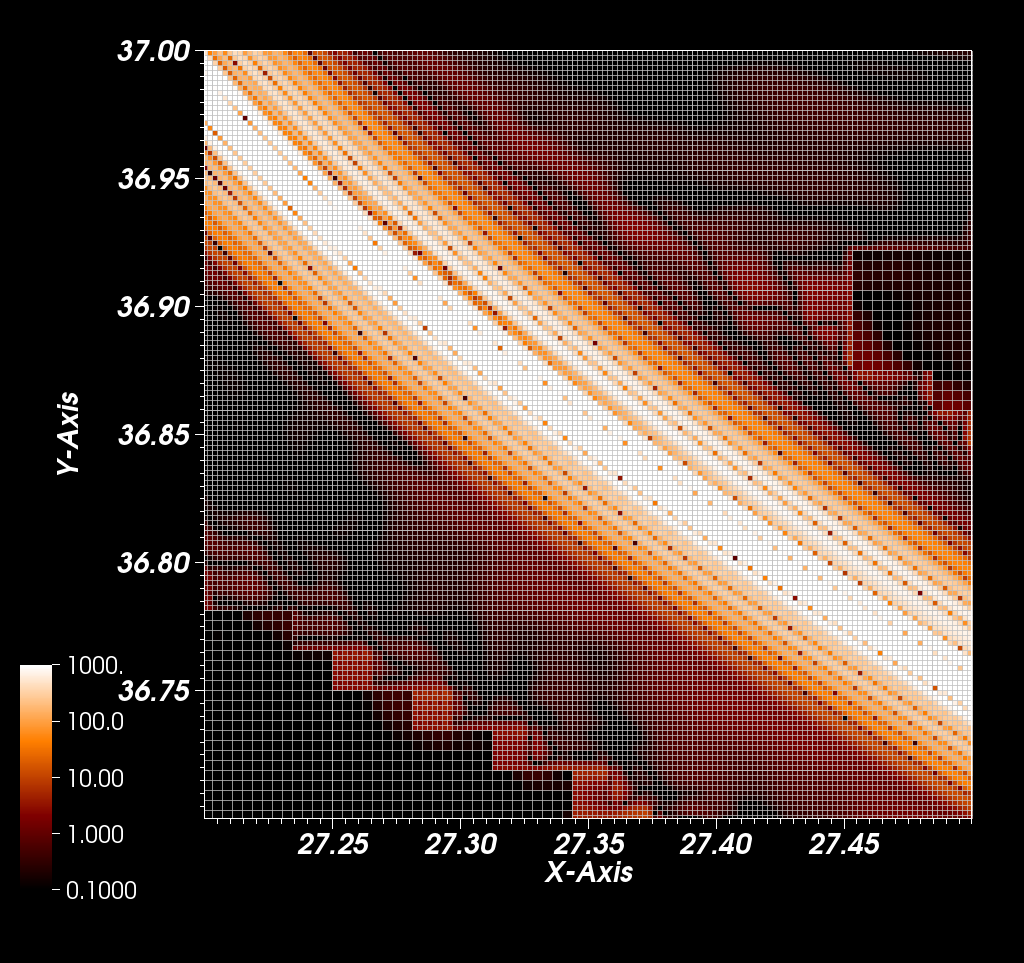}
  \caption{{\em Left column:} The Hamiltonian constraint for the
           6$D$-A simulation in the $xy$ plane at
           $t=22$. The upper panel shows the constraint obtained
           using 8 refinement levels only.
           The constraint with an additional 9th
           level is shown in the center panel and, amplified
           by a factor $Q_2=4$ for second-order convergence,
           in the bottom panel.
           {\em Right column:} A zoom-in of each of the left
           panels shows the constraint in a smaller region
           around the string. Here, we also display the
           grid structure in the form of a gray mesh.
           }
  \label{fig:conv1}
\end{figure*}

In all of our simulations, we ensure that the region contained within the $\chi=0.6$ contour is covered by at least 40 grid points. Unacceptably large constraint violations due to insufficient resolution lead to unphysical effects, such as a rapid swelling of the horizon.

In view of the very high computational resources needed for the simulations presented in this work (about 2M CPU hours per run), a standard
convergence analysis with two additional simulations at higher
resolution is prohibitively costly. In order to achieve a practical
convergence study, we apply two steps that greatly reduce the
expenses. First, we consider a quantity whose continuum limit
is already known, namely the Hamiltonian constraint. This reduces the
number of required resolutions from three to two. Second, instead
of repeating the entire simulation, we use the fact
that the code dynamically adds resolution in the critical
regions by adding extra refinement levels. More specifically,
we have performed the later stages of the 6$D$-A simulation twice.
In one case, the code adds, as instructed by the tagging
criterion, an additional refinement level, the 9th level,
that covers the black string region and the satellites. In the second case,
we have disallowed the code from adding this level and
instead let the simulation continue from its previous checkpoint
with a mere 8 refinement levels. These two instances of our
simulation enable us to analyse the dependence of the Hamiltonian
constraint on resolution in the vicinity of the high curvature region.
We note in this context that the constraint violations rapidly decrease
in the outer regions of our computational domain, as expected from
the rapid fall-off of gravity in higher-dimensional
spacetimes.

In Fig.~\ref{fig:conv1}, we show the Hamiltonian constraint
for the 6$D$-A simulation in the $xy$ plane at $t=22$.
For better image quality,
we display only the upper left half of the string
in the left column of this figure;
the lower right half is obtained from reflection
across each panel's bottom right corner. Let us first compare the upper and
center panels which are obtained using 8 and 9 levels, respectively.
We clearly see that the string shaped region of large
constraint violations is thicker for the lower resolution in
the upper panel. The resulting larger constraint
violations also extend into the outer regions as visible
in the form of the cloud like structure surrounding the
string, especially in the upper half. Both these effects
are visibly reduced by the higher resolution around the string
in the center panel, although at a small price: the additional
refinement boundaries introduce some numerical noise visible
in the form of the rectangular structures in the center panel.

In order to obtain a rough estimate of the convergence order,
we show in the bottom left panel of Fig.~\ref{fig:conv1} the
higher resolution result amplified by a factor $Q_2=4$ as
expected for second-order convergence, which is the dominant
discretization order in the prologation operation. This amplification
roughly widens the region of larger constraint violations
in the bottom panel to
a level comparable to the top panel. Note that we only
expect this convergence in the immediate vicinity of the
string, where extra refinement is present
in the center and bottom panels.

In order to obtain a clearer illustration of the rate of
convergence, we show in the right column of Fig.~\ref{fig:conv1}
the same result but now for a
small region around the string, where we are now also able
to display the grid structure. Again, the effect of
the additional resolution near the string in the lower
two panels is clearly perceptible. The comparable width
of the string in the top and bottom panel furthermore
indicates convergence commensurate with 2nd order around the string.

\bibliographystyle{JHEP-2}
\bibliography{refs}

\end{document}